\documentclass[12pt]{article}
\usepackage{graphicx}
\usepackage{amssymb}
\usepackage{amsmath}
\usepackage{color}

\newcommand{\bear}{\begin{array}}  
\newcommand {\eear}{\end{array}}
\newcommand{\bea}{\begin{eqnarray}}   
\newcommand{\eea}{\end{eqnarray}}
\newcommand{\beq}{\begin{equation}}   
\newcommand{\eeq}{\end{equation}}
\newcommand{\bef}{\begin{figure}}  \newcommand 
{\eef}{\end{figure}}
\newcommand{\bec}{\begin{center}}  \newcommand 
{\eec}{\end{center}}

\begin{document}

\begin{titlepage}

\begin{flushright}
ICRR-Report-542 \\
IPMU 09-0043 \\
\end{flushright}

\vskip 1cm

\begin{center}
{\large \bf
Diffuse gamma-ray background and cosmic-ray positrons\\
from annihilating dark matter 
}
\vskip 1cm

Masahiro Kawasaki$^{(a,b)}$,
Kazunori Kohri$^{(c)}$ and
Kazunori Nakayama$^{(a)}$
 
\vskip 0.5cm

{\it $^a$Institute for Cosmic Ray Research,
University of Tokyo, Kashiwa 277-8582, Japan}

\vskip 0.3cm

{\it $^b$Institute for the Physics and Mathematics of the Universe,
University of Tokyo, Kashiwa 277-8568, Japan}

\vskip 0.3cm

{\it $^c$Physics Department, Lancaster University, Lancaster LA1 4YB, UK}

\vskip 0.3cm

\begin{abstract}
We study the annihilating dark matter contribution to the
extra-galactic diffuse gamma-ray background spectrum,
motivated by the recent observations of cosmic-ray positron/electron anomalies.
The observed diffuse gamma-ray flux provides stringent constraint on
dark matter models and we present upper bounds on the annihilation cross section of the dark matter.
It is found that for the case of cored dark matter halo profile, 
the diffuse gamma-rays give more stringent bound compared with gamma-rays
from the Galactic center.
The Fermi satellite will make the bound stronger.
\end{abstract}

\end{center}
\end{titlepage}

%%%%%%%%%%%%%%%%%%%%%%%%%%%%%%%%%%%%
\section{Introduction}
%%%%%%%%%%%%%%%%%%%%%%%%%%%%%%%%%%%%

It is known that nearly 20\% of the total energy density of the present Universe
is dominated by the unknown non-relativistic matter, called dark matter (DM) \cite{Komatsu:2008hk}.
One of the main goals of particle physics is to reveal or determine the nature of DM,
since it is believed that DM is closely related to the physics beyond the standard model.
Some methods are proposed in order to approach the properties of DM \cite{Jungman:1995df}.
One is a direct detection experiment, which searches for DM scattering signals
with nucleons inside the detector.
Another way is to search for characteristic signatures of DM annihilation in the cosmic-ray flux,
such as gamma-rays, positrons/electrons, anti-protons and neutrinos.

Interestingly enough, one clue to the DM may have been found in cosmic-ray signals.
Recent observation of the PAMELA satellite found a steep excess in the cosmic positron flux
above the energy $\sim 10$~GeV~\cite{Adriani:2008zr}.
The ATIC balloon experiment also reported an excess in a sum of electron and positron flux
whose peak energy is around 600~GeV~\cite{:2008zz}. 
This steep excess was not confirmed by the Fermi satellite~\cite{Abdo:2009zk},
but still there may be a deviation from the astrophysical background electron flux.
These positron/electron excesses can be interpreted as a signal of DM annihilation,
and this possibility was discussed in many literature~\cite{DMannihilation}. 
In these scenarios, the annihilation cross section which is two or three orders of
magnitude larger than the canonical value 
($\langle \sigma v\rangle \sim 3\times 10^{-26}~{\rm cm^3s^{-1}}$),
which reproduces a correct relic DM abundance under thermal freeze out scenario, is required.
One way to incorporate this is to rely on unknown clumpy structure of DM halo
in the Galaxy.
If DM forms a small scale clumps, DM-originated positron/electron flux can be enhanced.
Such an enhancement is characterized by a so-called boost factor $B_F$,
and if $B_F$ reaches to $\mathcal{O}(10^2)-\mathcal{O}(10^3)$, 
the PAMELA/Fermi results can be reproduced within the canonical value of 
DM annihilation cross section.
However, one needs not introduce a large boost factor if one extends a DM generation mechanism
to a non-thermal production~\cite{Kawasaki:1995cy,Moroi:1999zb}.
In such cases, the DM annihilation cross section can take a large value
while accounting for the present DM abundance.
This latter possibility is interesting because indirect detection signatures are also 
generically enhanced~\cite{Profumo:2004ty}.

Motivated by the recent PAMELA/Fermi observations,
we investigate gamma-ray signatures from DM annihilation scenario.
In particular, the study of diffuse extragalactic gamma-ray background
in the context of recent anomalous positron flux was missed in the literature.
At first sight, DM annihilation may not seem to contribute significantly to
diffuse extragalactic component because gamma-ray flux from DM annihilation is proportional to
the square of DM density, and DM density in intergalactic space is so low.
However, taking into account DM annihilation inside the external galaxies 
and clumpy objects leads to a great enhancement, 
which can compensate the quite dilute DM density 
\cite{Bergstrom:2001jj,Ullio:2002pj} 
(see also Refs.~\cite{Taylor:2002zd,Ahn:2004yd,Oda:2005nv}).
We find that diffuse extragalactic gamma-ray background gives a stringent constraint
on DM annihilation models which account for the PAMELA/Fermi anomaly.
The forthcoming Fermi satellite experiment will soon provide more useful constraints 
in a similar way \cite{Baltz:2008wd}, 
and hence it is important to reconsider diffuse gamma-ray bounds on DM annihilation models.
We have improved the previous estimations taking into account following points.
First, we adopt WMAP5 best fit cosmological parameters.
Also we used an improved transfer function by Eisenstein and Hu \cite{Eisenstein:1997jh}, 
rather than conventional BBKS one \cite{Bardeen:1985tr}.
Finally, we calculate the gamma-ray spectrum for various final states including
both Galactic component and extragalactic component.
As for the extragalactic component, gamma-rays from homogeneously distributed DM
are also included.
Each of these improvements yields more than factor of two or three modification
on the resultant gamma-ray spectrum.

This paper is organized as follows.
First in Sec.~\ref{sec:pos} we present the cosmic-ray positron flux from DM annihilation
in order to provide representative DM models reproducing PAMELA anomaly.
In Sec.~\ref{sec:GC} we first evaluate the gamma-ray flux from the Galactic center.
Then in Sec.~\ref{sec:extra} we study the diffuse extragalactic gamma-rays 
and derive bounds on the annihilation rate.
Sec.~\ref{sec:conc} is devoted to conclusions and discussion.

%%%%%%%%%%%%%%%%%%%%%%%%%%%%%%%%%%%%%%%%%
\section{Cosmic positron signature from dark matter annihilation} \label{sec:pos}
%%%%%%%%%%%%%%%%%%%%%%%%%%%%%%%%%%%%%%%%%

Propagations of cosmic ray positrons/electrons produced by DM annihilation in the Galactic halo
are described by a random walk process because of the effect of tangled magnetic field
in the Galaxy, with losing its energy due to the synchrotron emission and inverse-Compton scattering
off microwave background photon and diffuse star light \cite{Baltz:1998xv}.
To calculate the electron and positron fluxes from the DM
annihilation, we have followed a conventional procedure \cite{Hisano:2005ec}.  
Approximating the shape of the diffusion zone
as a cylinder (with half-height $L$ and the length $2R=40~{\rm kpc}$), 
the static solution to the following diffusion equation are derived:
\begin{equation}
   \frac{\partial}{\partial t}f(E, \vec x) = K(E)\nabla^2f(E, \vec x)
   +\frac{\partial}{\partial E} [b(E)f(E, \vec x)] + Q(E,\vec x), 
   \label{diffusion}
\end{equation}
where $K(E)$ is the diffusion constant, $Q(E,\vec x)$ is the source term from the
DM annihilation $b(E)=1\times 10^{-16}(E/1~{\rm GeV})^2$ is the energy loss rate
and $f(E, \vec x)$ is the electron/positron number density with
energy $E$ at the position $\vec x$, which is related to the DM
contribution to the electron and positron fluxes as
\begin{equation}
   \frac{d\Phi^{(\rm DM)}_{e^+}}{dE}(\vec x_\odot)
   = \frac{c}{4\pi}f(E,\vec x_\odot),
\end{equation}
where $c$ is the speed of light and $\vec x_\odot$ denotes the
position of the solar system. 
As for the half height of the diffusion zone $L$ and the diffusion constant
parameterized by $K(E)=K_0(E/1~{\rm GeV}^\delta)$,
we adopt so-called MED and M2 propagation models \cite{Delahaye:2007fr} :
$(K_0, \delta, L)=(0.0112~{\rm kpc^2/Myr}, 0.70, 4~{\rm kpc})$ for MED model,
and $(K_0, \delta, L)=(0.00595~{\rm kpc^2/Myr}, 0.55, 1~{\rm kpc})$ for M2 model, 
both of which are consistent with observed boron-to-carbon ratio in the cosmic-ray flux.

The source term in Eq.~(\ref{diffusion}) is given by
\begin{equation}
  Q(E,\vec x) = \frac{1}{2} B_F\left( \frac{\rho_{\rm gal}(\vec x)}{m_\chi}\right)^2 
  \sum _F \left (\langle \sigma v \rangle_F
  \frac{dN^{(e^+)}_F}{dE} \right ),
  \label{sourceterm}
\end{equation}
where $\rho_{\rm gal}(\vec x)$ is the present mass distribution of DM in the Galactic halo.,
$m_\chi$ is the DM mass, $F$ collectively denotes the final state of DM annihilation,
$\langle \sigma v \rangle_{F}$ denotes the annihilation cross section into the final state $F$.
The fragmentation function, $dN_{F}^{(e^+)}/dE$, represents the spectrum of 
positrons generated per DM annihilation into the final state $F$,
evaluated by PYTHIA package~\cite{Sjostrand:2006za}.
DM may not be distributed smoothly in the Galaxy and can have clumpy structure
within the region 1~kpc around the solar system.
If this is the case, the positron flux is enhanced and such an effect is represented by 
the boost factor, $B_F$. In the following we conservatively assume $B_F=1$.
It should be noted that the unknown clumpy structure of DM halo always
introduces an uncertainty on the annihilation cross section to reproduce the PAMELA anomaly.
On the other hand,
dependence of the electron and positron fluxes on the DM density profile is very minor 
because typical propagation distance of the positrons/electrons with energy of our interest is
less than 1~kpc.

\begin{figure}[t]
 \begin{center}
 \includegraphics[width=0.45\linewidth]{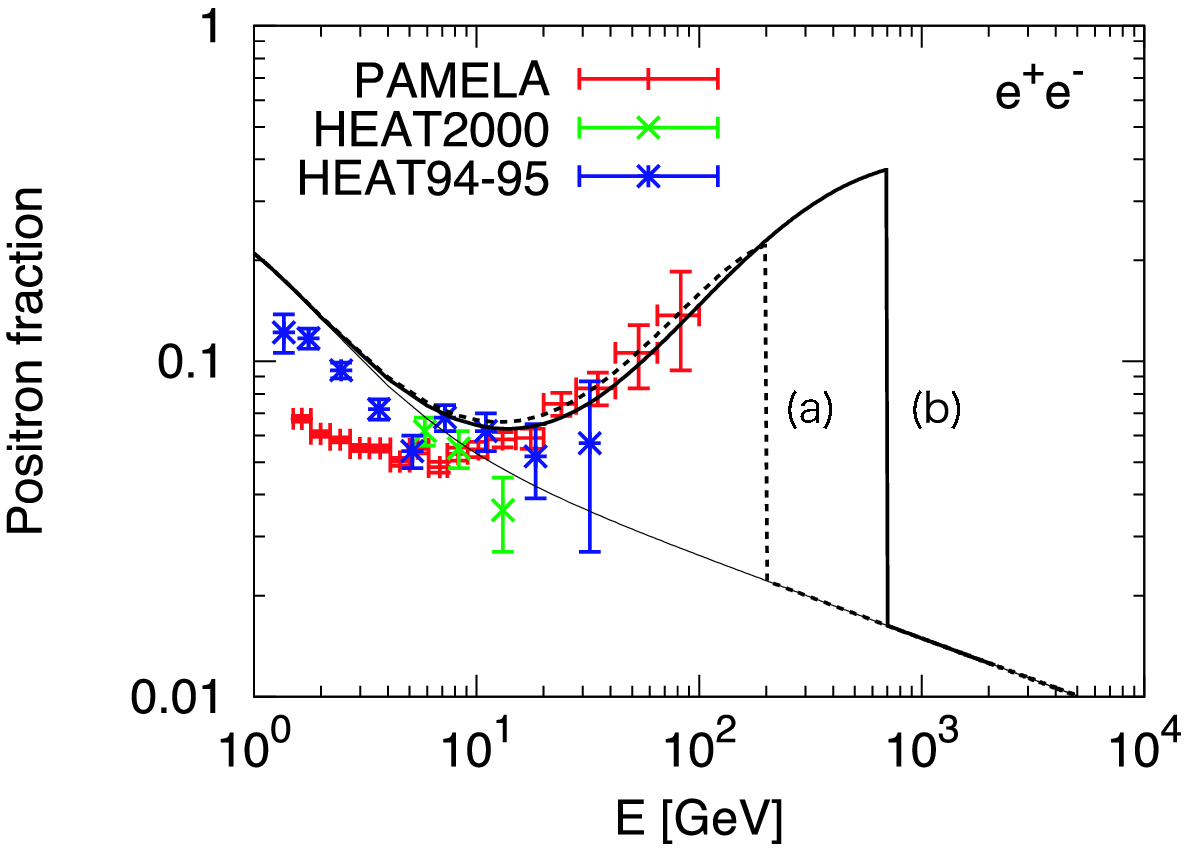} 
 \includegraphics[width=0.45\linewidth]{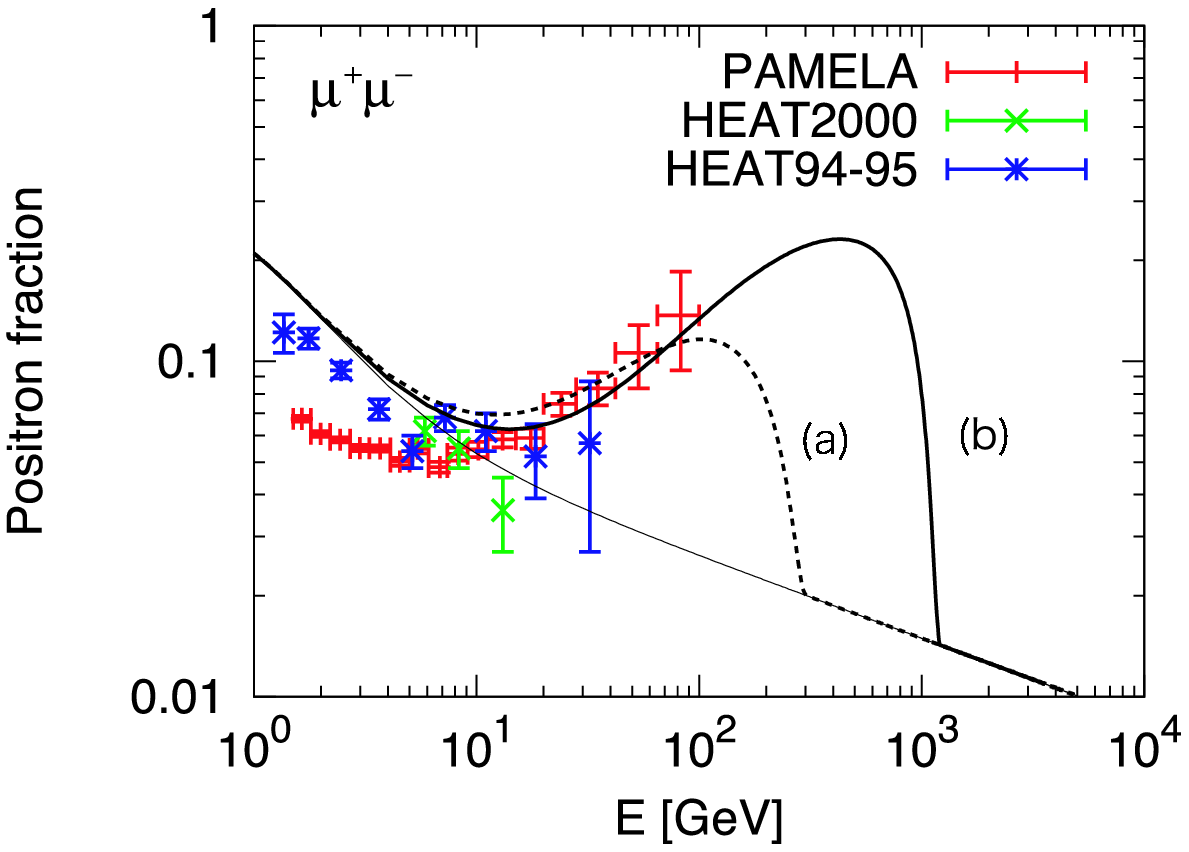} 
 \includegraphics[width=0.45\linewidth]{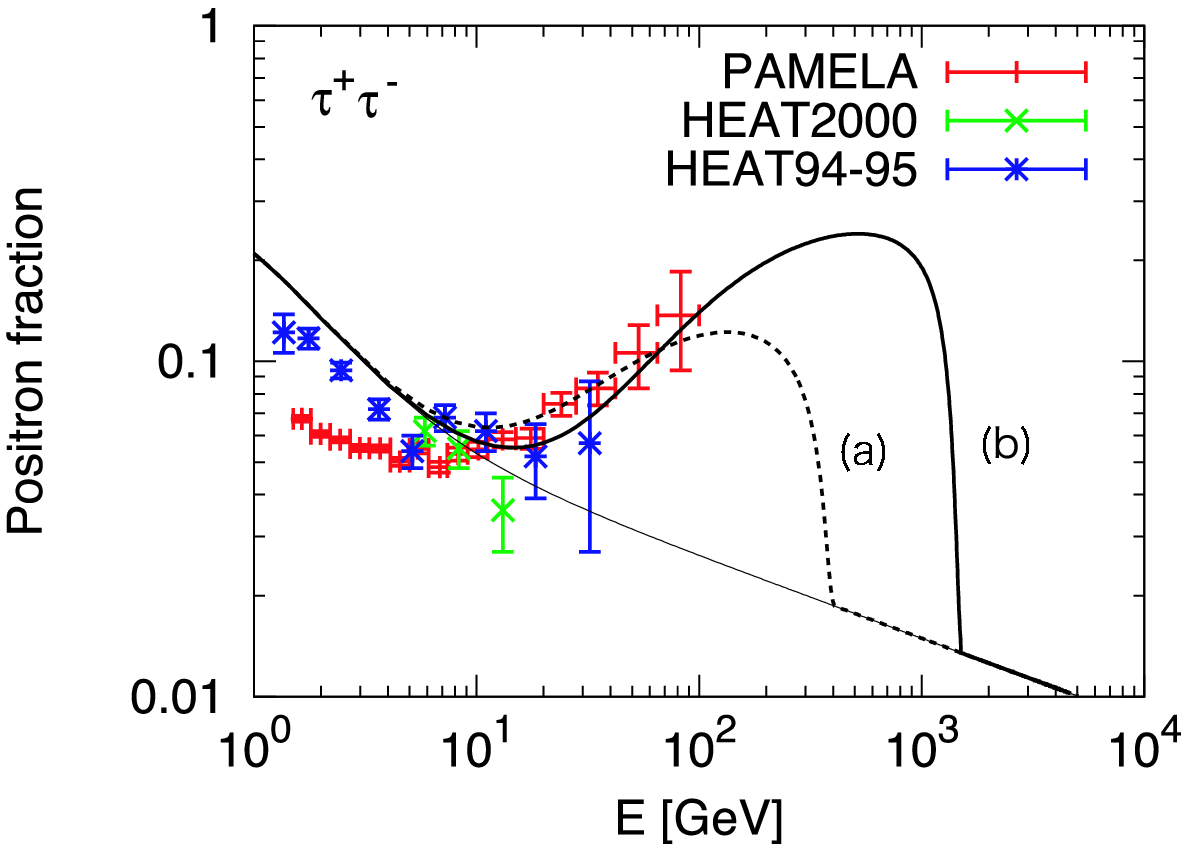} 
 \includegraphics[width=0.45\linewidth]{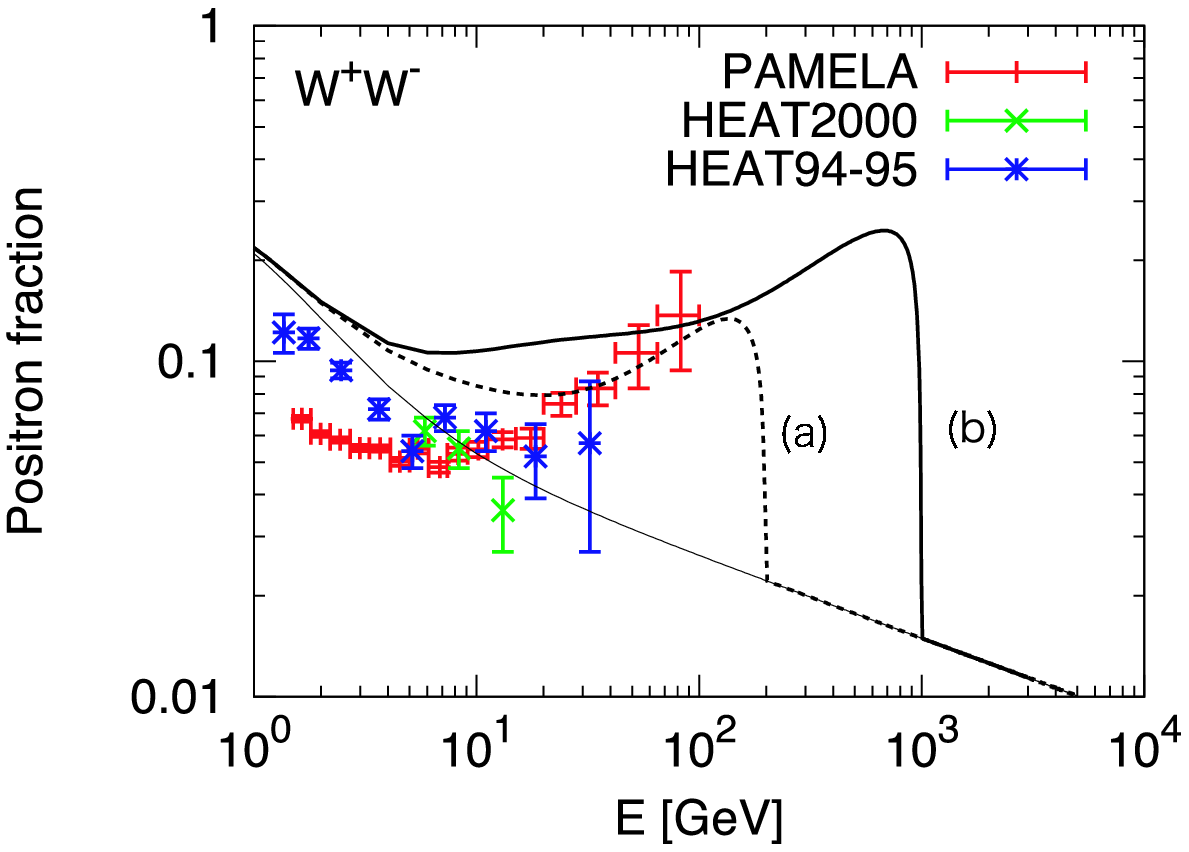} 
 \caption{ Positron flux for DM models (a) and (b) for various final states
 given in table.~\ref{table:model}.  }
  \label{fig:pos}
 \end{center}
\end{figure}

Fig.~\ref{fig:pos} shows resultant positron flux, in terms of the positron fraction
defined by the ratio of the positron flux to the sum of electron and positron flux, are shown.
The background flux is taken from Ref.~\cite{Moskalenko:1997gh}.
The PAMELA~\cite{Adriani:2008zr} and HEAT~\cite{Barwick:1997ig} results are shown.
As for DM models, we consider two representative parameter sets :
model (a) for rather light DM mass which explains the PAMELA anomaly, and
model (b) for heavier DM which can explain both the PAMELA and Fermi observations.
DM masses and annihilation cross sections for various final states
are summarized in table~\ref{table:model}.
We adopt MED propagation model for the DM annihilating into $e^+e^-$ and $\mu^+\mu^-$,
and M2 model into $\tau^+\tau^-$ and $W^+W^-$.
Both models (a) and (b) well reproduce recent PAMELA anomaly and
the Fermi electron excess can also be explained for model (b).

%%%%%%%%%%%%%%%%%%FIGURE%%%%%%%%%%%%%%%%%%%
%%%%%%%%%%%%%%%%%%%%%%%%%%%%%%%%%%%%%%%%%

%%%%%%%%%%%%%%%%  table %%%%%%%%%%%%%%%%%%%%%%
\begin{table}[ht]
  \begin{center}
    \begin{tabular}{l  |  lll  |  lll }
      \hline \hline
      Final State 
      & (a)
      & $m_{\rm DM}\ [{\rm GeV}]$
      & $\langle\sigma v\rangle\ [{\rm cm^3s^{-1}}]$
      & (b)
      & $m_{\rm DM}\ [{\rm GeV}]$
      & $\langle\sigma v\rangle\ [{\rm cm^3s^{-1}}]$   \\
      \hline
      $e^+ e^-$             &  & 200  & $0.4\times 10^{-24}$ & & 700  & $0.4\times 10^{-23}$ \\
      $\mu^+ \mu^- $ &  & 300  & $1.2\times 10^{-24}$ & & 1200  & $1.2\times 10^{-23}$  \\
      $\tau^+ \tau^- $ &  & 400  & $5.0\times 10^{-24}$ & & 1500 & $3.5\times 10^{-23}$ \\
      $W^+ W^-$           &  & 200  & $4.0\times 10^{-24}$ & & 1000  & $3.0\times 10^{-23}$ \\
      \hline \hline
    \end{tabular}
    \caption{The DM particle mass and annihilation cross section used in Fig.~\ref{fig:pos}.}
    \label{table:model}
  \end{center}
\end{table}
%%%%%%%%%%%%%%%%%%%%%%%%%%%%%%%%%%%%%%%%%%%%%% 

%%%%%%%%%%%%%%%%%%%%%%%%%%%%%%%%%%%%%%%%%
\section{Dark matter contribution to gamma-rays from Galactic center} \label{sec:GC}
%%%%%%%%%%%%%%%%%%%%%%%%%%%%%%%%%%%%%%%%%

The DM density strongly enhances at the Galactic center (GC), and hence 
the gamma-ray flux coming from the DM annihilation is expected to be largely enhanced,
depending on the DM halo density profile.
The gamma-ray flux from the GC is expressed as \cite{Bergstrom:1997fj}
\begin{equation}
	\left[\frac{d\Phi_\gamma(\psi)}{dE}\right]_{\rm Gal}
	 = \sum_{F} \frac{\langle \sigma v \rangle_F}{8\pi m_\chi^2}
	\frac{dN_{F}^{(\gamma)}}{dE}
	\int_{\rm l.o.s.}\rho_{\rm gal}^2(l)dl(\psi),   \label{fluxGCpsi}
\end{equation}
where $\rho_{\rm gal}$ is the DM density in the Galaxy,
$l(\psi)$ denotes the distance from the Earth along the direction $\psi$.
The fragmentation function, $dN_{F}^{(\gamma)}/dE$, represents the spectrum of 
gamma-rays generated per DM annihilation into the final state $F$.
The integration is performed along the line of sight toward the direction $\psi$.

Gamma-rays are emitted by DM annihilation
mainly by two processes : internal bremmstrahlung and cascade decay.
In the case where the final state particles have electric charge, 
as is the case of most realistic DM models including SUSY neutralinos,
the final state particles can always emit gamma-rays with branching ratio suppressed by 
the electromagnetic fine structure constant.
To avoid complexity and to perform a model independent analysis, 
we only consider the final state radiation as internal bremmstrahlung processes,
although there may be other important contributions from initial state radiation and/or
virtual internal bremmstrahlung \cite{Bergstrom:2005ss}.
Cascade decay of final state particles also produce significant amount of gamma-rays,
mainly coming from the $\pi^0$ decay, if the final states have hadronic decay modes.
This is evaluated by PYTHIA package \cite{Sjostrand:2006za}.

Actual observations have finite angular resolution,
and hence what we can compare with observations is gamma-ray flux integrated over some region.
Integrating (\ref{fluxGCpsi}) over the solid angle $\Delta \Omega$ around the GC,
we obtain the gamma-ray flux
\begin{equation}
	\left[\frac{d\Phi_\gamma}{dE}\right]_{\rm Gal} 
	= \frac{R_\odot \rho_\odot^2}{8\pi m_\chi^2}
	\left (\sum_F \langle \sigma v \rangle_F \frac{dN_{F}^{(\gamma)}}{dE} \right )
	\langle J \rangle_\Omega \Delta \Omega,
\end{equation}
where $R_\odot=8.5$~kpc and $\rho_\odot=0.3~{\rm GeV~cm^{-3}}$ are
the distance from the GC and local DM density near the solar system, and
the dependence on the DM halo profile is contained in the $J$-factor,
\begin{equation}
	\langle J \rangle_\Omega = \int \frac{d\Omega}{\Delta \Omega}
	\int_{\rm l.o.s.} \frac{dl(\psi)}{R_\odot}\left( \frac{\rho_{\rm gal}(l)}{\rho_\odot}\right )^2.
\end{equation}
The integration over the solid angle is performed within $0<\psi<\psi_{\rm max}$,
with $\Delta \Omega=2\pi(1-\cos \psi_{\rm max})$.
We choose $\psi_{\rm max}=0.1^\circ$, corresponding to $\Delta \Omega=10^{-5}~{\rm sr}$,
taking into account the angular resolution of ground based Cherenkov telescope such as HESS.

The $J$-factor sensitively depends on DM halo profile models.
We characterize it by two parameters, a typical density scale $\tilde \rho$ and radial scale $a$, 
as $\rho_{\rm gal}(r)=\tilde \rho d(r/a)$.
We consider the following halo models : 
(a) Navarro-Frenk-White (NFW) profile \cite{Navarro:1995iw}, 
\begin{equation}
	 d_{\rm NFW}(x)=\frac{1}{x(1+x)^2}, 
\end{equation}
(b) Moore profile \cite{Moore:1999gc},
\begin{equation}
	 d_{\rm Moore}(x)=\frac{1}{x^{3/2}(1+x^{3/2})}, 
\end{equation}
(c) Burkert profile \cite{Burkert:1995yz}
\begin{equation}
	 d_{\rm Bur}(x)=\frac{1}{(1+x)(1+x^2)}.
\end{equation}
While NFW and Moore profiles have cuspy structure at $r \to 0$,
all of them show a behavior like $\rho(r) \sim r^{-3}$ in the large $r$ limit.

\begin{figure}[ht]
 \begin{center}
 \includegraphics[width=0.45\linewidth]{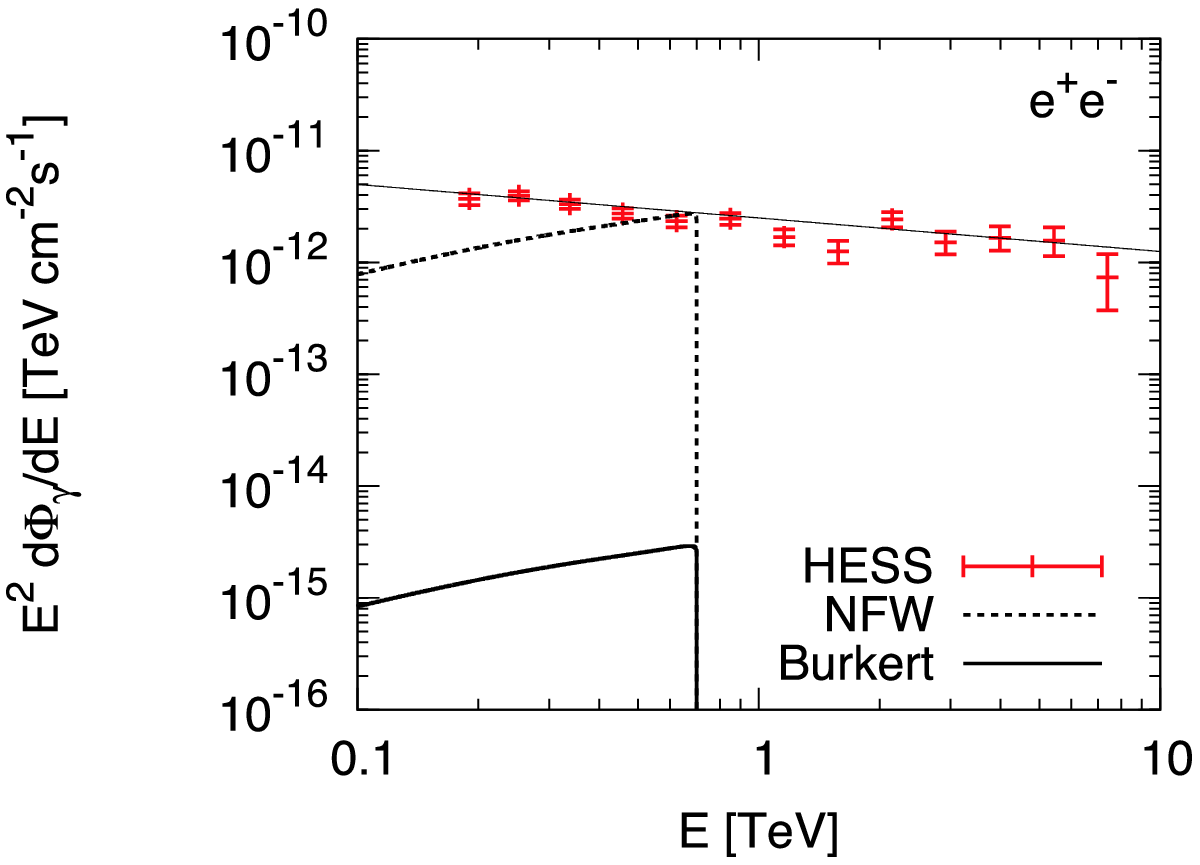} 
 \includegraphics[width=0.45\linewidth]{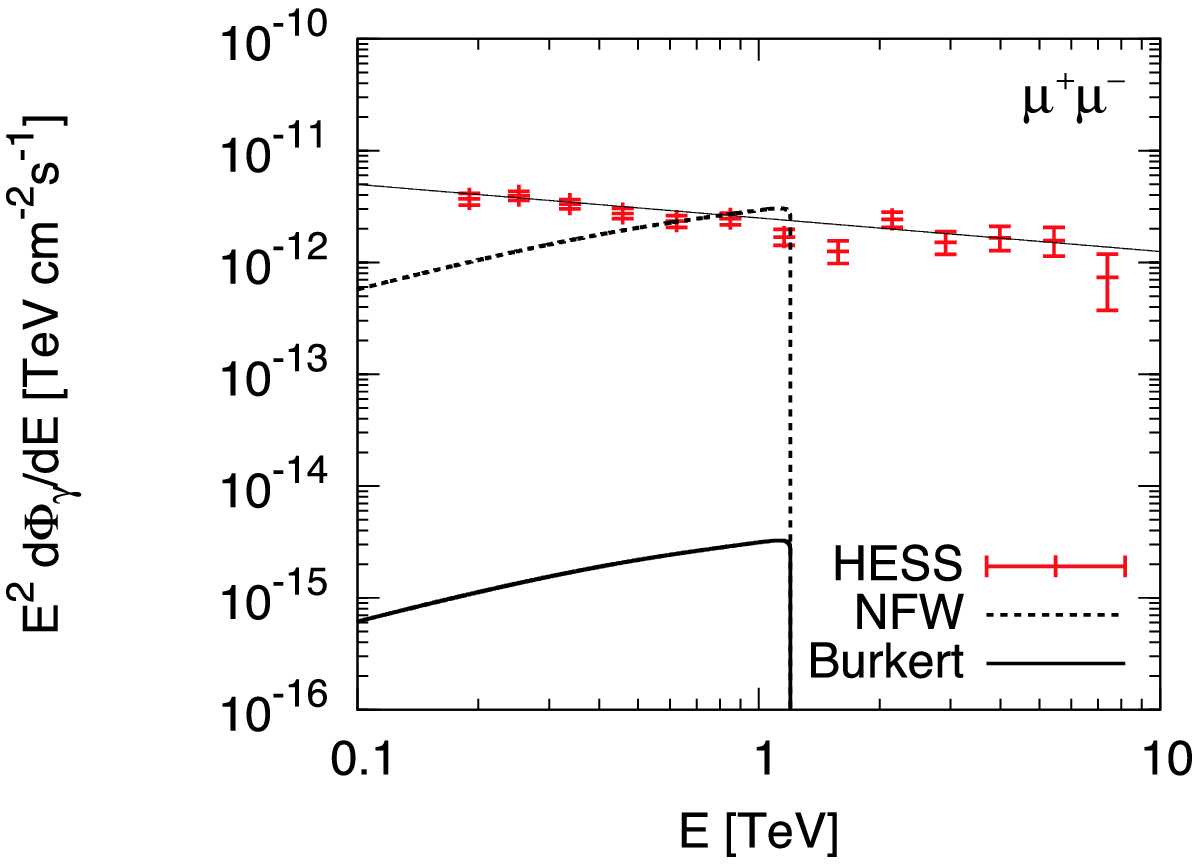} 
 \includegraphics[width=0.45\linewidth]{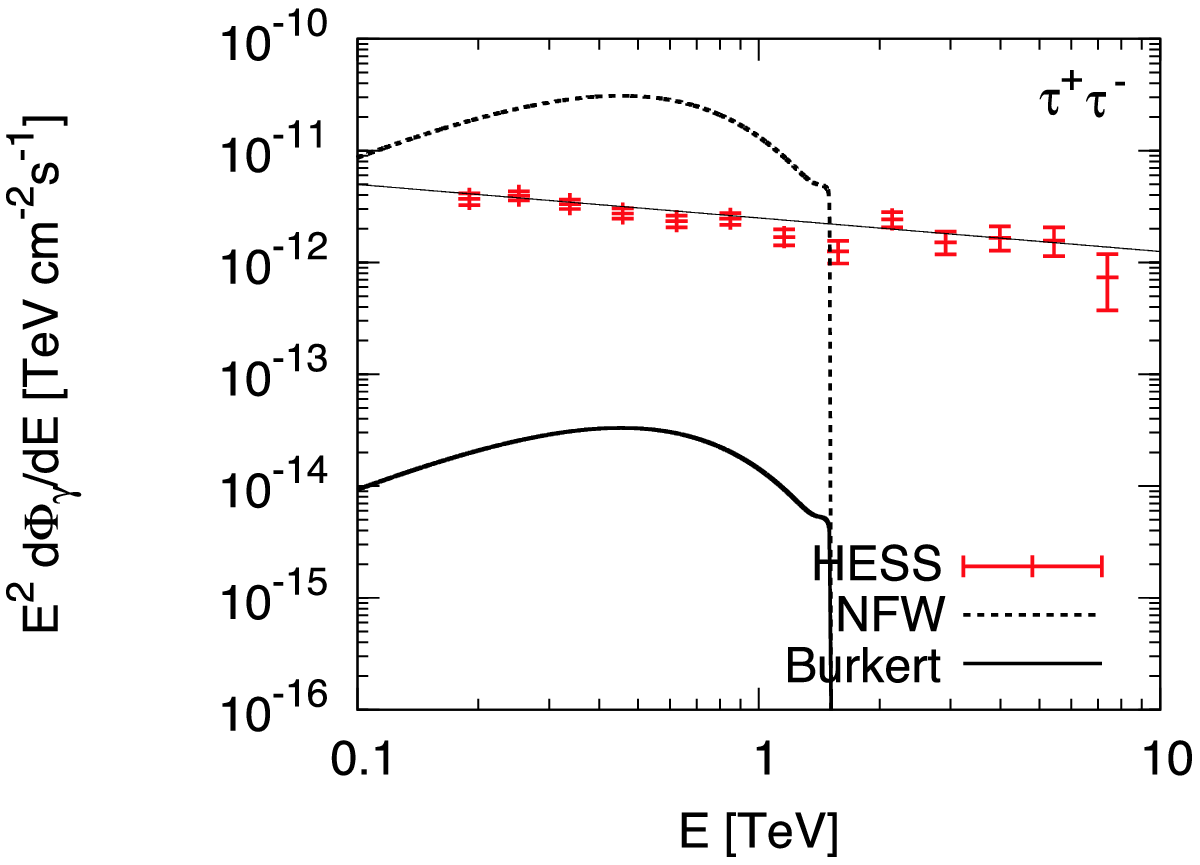} 
 \includegraphics[width=0.45\linewidth]{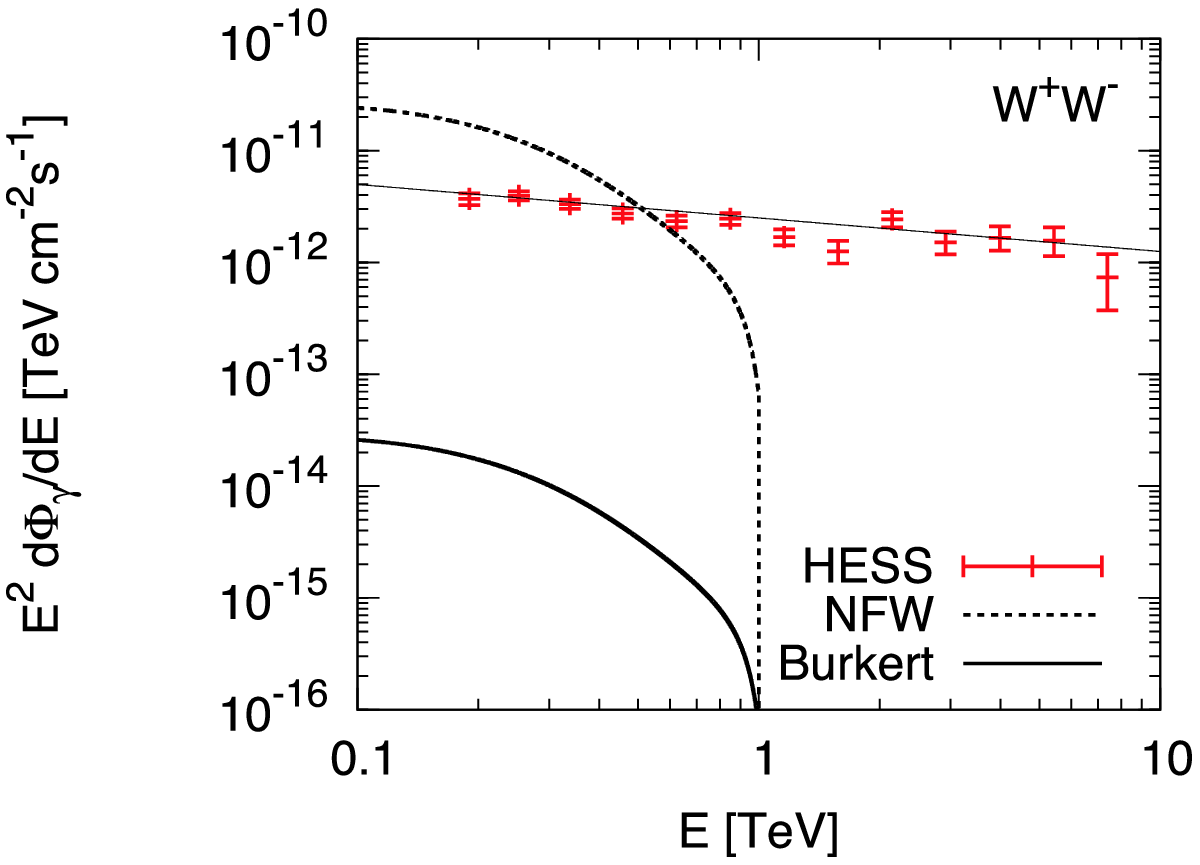} 
 \caption{ DM annihilation contribution to the gamma-ray flux from the GC with solid angle
  $\Delta \Omega=10^{-5}~{\rm sr}$. 
  We adopt DM models in table.~\ref{table:model} : DM annihilating into 
  $e^+e^-$ (top left), $\mu^+\mu^-$ (top right), $\tau^+\tau^-$ (bottom left) and $W^+W^-$ (bottom right). }
  \label{fig:GC}
 \end{center}
\end{figure}

Fig.~\ref{fig:GC} shows the gamma-ray flux from the GC with the HESS results \cite{Aharonian:2004wa}.
For reference, we depicted a line which fits the HESS results as
$d\Phi_\gamma/dE = 2.5\times 10^{-12}(E/{\rm TeV})^{-2.3}$\\
$[{\rm TeV^{-1}cm^{-2}s^{-1}}]$.
We adopt four DM models :
DM annihilating into $e^+e^-$, $\mu^+\mu^-$, $\tau^+\tau^-$ and $W^+W^-$.
As for the halo profile, we adopt NFW and Burkert models.
Moore profile predicts about one order of magnitude larger flux than the NFW profile,
and hence is not favored.
As can be seen in the figure, a cuspy DM density profile such as NFW profile
would produce too large gamma-ray flux, which exceeds the HESS observations of the GC.
However, if the DM halo has a core, as in the Burkert profile,
the gamma-ray flux is significantly reduced, so that it can be compatible with the HESS observation.
These results are consistent with e.g. Ref.~\cite{Bertone:2008xr}.

%%%%%%%%%%%%%%%%%%FIGURE%%%%%%%%%%%%%%%%%%%
%%%%%%%%%%%%%%%%%%%%%%%%%%%%%%%%%%%%%%%%%

%%%%%%%%%%%%%%%%%%%%%%%%%%%%%%%%%%%%%%%%%
\section{Dark matter contribution to diffuse extragalactic gamma-ray background}
\label{sec:extra}
%%%%%%%%%%%%%%%%%%%%%%%%%%%%%%%%%%%%%%%%%

In this section we evaluate the DM annihilation contribution to the diffuse extragalactic
gamma-ray background along the line of Ref.~\cite{Ullio:2002pj}.

As the Universe expands, structure formation due to gravitational collapses proceeds
and non-relativistic matter forms more and more halos with various scales.
Inside the halo, the DM annihilation rate is enhanced.
Thus we must take into account distributions of such DM halos and their effects
on the diffuse gamma-ray flux.
We write the gamma-ray flux from the halo with mass $M$ at redshift $z$ as
$d\phi_\gamma/dE'(M,z)$ where $E'$ is the gamma-ray energy at the production,
and is observed now as a photon with energy $E=E'/(1+z)$ due to the redshift.
Hereafter, we neglect the effect of intergalactic absorption of gamma-rays,
which is irrelevant for the energy range we are interested in.
The resulting diffuse gamma-ray flux is given by
\begin{equation}
	\left[\frac{d\Phi_\gamma}{dE}\right]_{\rm ext}=\frac{c}{4\pi}\int \frac{dz}{H_0 h(z)}
	\int dM \frac{dn(z)}{dM} \frac{d\phi_\gamma}{dE'}(M,z),
\end{equation}
where $c$ is the speed of light, $H_0$ is the present Hubble parameter,
$h(z)=\sqrt{\Omega_\Lambda+\Omega_m (1+z)^3}$,
$dn/dM(z)$ denotes the comoving number density
of halos with mass $M$ at redshift $z$ evaluated later.
In the following we evaluate the source gamma-ray flux $d\phi_\gamma/dE'(M,z)$
and distributions of DM halos $dn/dM(z)$.

%%%%%%%%%%%%%%%%%%%%%%%%%
\subsection{Profile of dark matter halos}
%%%%%%%%%%%%%%%%%%%%%%%%%

The DM annihilation rate from the halo with mass $M$ strongly depends on
the density profile of the halo.
$N$-body simulations indicate that the density profile of DM halos take universal form irrespective of 
the size of the halos.
We consider halo models described in the previous section : NFW, Moore and Burkert.
For later convenience, we define $I_n(x)$ as
\begin{equation}
	I_n(x_{\rm max})=\int^{x_{\rm max}}_{x_{\rm min}} dx x^2 d^n(x).  \label{I_n}
\end{equation}

Then we define virial radius $R_{\rm vir}(M)$
as a radius within which the mean background density times $\Delta_{\rm vir}$ becomes $M$,
\begin{equation}
	\frac{4\pi}{3}R_{\rm vir}^3(M)\Delta_{\rm vir}(z)\bar \rho_m(z) = M.
\end{equation}
Here the virial overdensity $\Delta_{\rm vir}(z)$ is given by \cite{Bryan:1997dn}
\begin{equation}
	\Delta_{\rm vir}(z) = \frac{18\pi^2+82y-39y^2}{\Omega_m(z)},
\end{equation}
where $\Omega_m(z)=\Omega_m(1+z)^3/h(z)^2$ and $y=\Omega_m(z)-1$.
Then we introduce a concentration parameter $C_{\rm vir}$ which is defined by
$C_{\rm vir}\equiv R_{\rm vir}/r_{-2}$, where $r_{-2}$ corresponds to the radius 
at which the slope of the density profile becomes $-2$ : 
a radius such that $d/dr(r^2\rho(r))=0$.
For NFW profile, $r_{-2}$ coincides with $a$.
It should be noticed that there is a non-trivial relation between 
the virial radius $R_{\rm vir}$ (or halo mass $M$)
and concentration parameter $C_{\rm vir}$ (or typical halo size $a$)
at each given redshift $z$~\cite{Bullock:1999he}.
Then we assume a log-normal distribution of the concentration parameter $C_{\rm vir}$,
expressed as $P(C_{\rm vir})$.

As a result, the gamma-ray spectrum from a single halo with mass $M$ is given by
\begin{equation}
\begin{split}
	 \frac{d\phi_\gamma}{dE'}(M,z)&=\sum_F\frac{\langle \sigma v \rangle _F}{2m_\chi^2}
	 \frac{dN_F^{(\gamma)}}{dE'}
	 \int dC_{\rm vir }(M,z)P(C_{\rm vir}(M,z)) \int d^3r \rho^2(r) \\
	 & =\sum_F\frac{\langle \sigma v \rangle _F}{2m_\chi^2}
	  \frac{dN_F^{(\gamma)}}{dE'} \frac{\bar \rho_m(z) M\Delta_{\rm vir}(z)}{3}
	  \int dC_{\rm vir }(M,z)P(C_{\rm vir}(M,z)) 
	  \frac{I_2(x_{\rm max})}{[I_1(x_{\rm max})]^2}x_{\rm max}^3,
\end{split}
\end{equation}
where $x_{\rm max}=R_{\rm vir}/a=C_{\rm vir}r_{-2}/a$ and
$I_n(x_{\rm max})~(n=1,2)$ are defined in Eq.~(\ref{I_n}).
Here we have used the relation $\int ^{R_{\rm vir}}\rho(r)d^3r = M$.
Then we must add up contributions from all DM halos and all redshift.
In order to do so, the mass distribution of DM halo must be known, and we evaluate it below.

%%%%%%%%%%%%%%%%%%%%%%%%%
\subsection{Distribution of dark matter halos}
%%%%%%%%%%%%%%%%%%%%%%%%%

According to the Press-Schechter theory~\cite{Press:1973iz}, the number of object with mass $M$
is expressed as
\begin{equation}
	\frac{dn(z)}{dM} = \sqrt \frac{2}{\pi} \frac{\bar \rho_m \delta_c(z)}{M}
	\left [ -\frac{1}{\sigma^2(M)}\frac{d\sigma(M)}{dM} \right ]
	\exp \left[ -\frac{\delta_c^2(z)}{2\sigma^2(M)} \right],  \label{PS}
\end{equation} 
where $\bar \rho_m$ is the average matter density,
$\delta_c(z)$ is the critical overdensity above which the spherical collapse occurs,
and $\sigma(M)$ is the variance of the density fluctuation containing the mass $M$,
whose definition is given below.
Using the growth factor $D_1(z)$, defined as a growth of the density perturbation 
inside the horizon after matter-radiation equality era, 
the critical overdensity can be written as $\delta_c(z)=1.67[D_1(z=0)/D_1(z)]$~\cite{Eke:1996ds}.
%In the limit of matter dominated Universe, it is given by $D_1(z)=(1+z)^{-1}$,
%but the presence of cosmological constant modifies its form \cite{Eisenstein:1997jh}.
Note that all the matter are involved in some clumped objects in the Press-Schechter theory, 
as is clearly seen by the relation $\int dM M (dn/dM) =\bar \rho_m$.
In Ref.~\cite{Sheth:1999su} a new formula is provided taking account of an effect of
ellipsoidal collapse in order to improve a fit to $N$-body simulation, 
\begin{equation}
	\frac{dn(z)}{dM} = A\sqrt b\sqrt \frac{2}{\pi} \frac{\bar \rho_m \delta_c(z)}{M}
	\left [ -\frac{1}{\sigma^2(M)}\frac{d\sigma(M)}{dM} \right ]
	\left ( 1+\frac{\sigma (M)^{2q}}{a^q \delta_c(z)^{2q}} \right )
	\exp \left[ -\frac{\delta_c^2(z)}{2\sigma^2(M)} \right],
\end{equation}
where $b = 0.707$ and $q=0.3$ are dimensionless parameters
and $A$ is an overall normalization constant.
$\delta_c(z)$ is same as that of the spherical case.

$\sigma (M)$ is calculated from the matter power spectrum $P_\delta (k)$ as
\begin{equation}
	\sigma (M)^2 =\frac{1}{2\pi^2} \int \tilde W^2(kR_M) P_\delta(k) k^2 dk,
\end{equation}
where $\tilde W(x)=3(\sin x-x\cos x)/x^3$ is a top-hat window function,
which has an effect to smooth out fluctuation with scale smaller than $x$,
and $R_M$ is a radius within which a mass $M$ is contained under the uniform matter density,
defined by $(4\pi/3)R_M^3 \bar \rho_m = M$.
It is estimated as
\begin{equation}
	R_M = 9.5\times 10^{-2}~{\rm kpc}~(\Omega_m h^2)^{-1/3}
	\left ( \frac{M}{M_\odot} \right )^{1/3}.
\end{equation}
The matter power spectrum is expressed as
\begin{equation}
	P_\delta(k)=\frac{4}{25}\frac{2\pi^2}{k^3}\left ( \frac{k^2}{\Omega_m H_0^2} \right )^2
	\Delta_\zeta^2 (k) T^2(k) D_1^2(z=0),
\end{equation}
where $\Delta_\zeta^2 (k)=\Delta_\zeta^2 (k_*)(k/k_*)^{n_s-1}$
with WMAP normalization $\Delta_\zeta^2 (k_*)=2.457\times 10^{-9}$ at $k_*=0.002~{\rm Mpc}^{-1}$
and scalar spectral index $n_s$.
We adopt the transfer function $T(k)$ and the growth factor $D_1(z)$
given in Ref.~\cite{Eisenstein:1997jh}
with WMAP5 best fit cosmological parameters : $\Omega_\Lambda = 0.72, \Omega_m=0.28,
\Omega_b=0.046,h=0.70, n_s=0.96$.
Using these parameters, we obtain $D_1(z=0)=0.77$.
As a cross check, this yields $\sigma_8 \equiv \sigma(8h^{-1}{\rm Mpc})=0.82$,
which is consistent with WMAP5 result \cite{Komatsu:2008hk}.

%%%%%%%%%%%%%%%%%%%%%%%%%
\subsection{Diffuse gamma-ray flux}
%%%%%%%%%%%%%%%%%%%%%%%%%

Combining obtained results, we arrive at the following formula 
for the diffuse gamma-rays from DM annihilation:
\begin{equation}
	\left[\frac{d\Phi_\gamma}{dE}\right]_{\rm ext} = 
	\sum_F \frac{\langle \sigma v \rangle_F }{8\pi}\frac{\bar \rho_m^2}{m_\chi^2}
	\int dz (1+z)^3 \frac{dN_F^{(\gamma)}}{dE'} \frac{\Delta^2(z)}{H_0 h(z)},  \label{ext}
\end{equation}
where the enhancement factor $\Delta^2(z)$ is given by
\begin{equation}
	\Delta^2(z) = \frac{1}{\bar \rho_m}\int_{M_{\rm cut}} dM M \frac{dn(z)}{dM}\Delta_M^2(M,z)
	+ f_{\rm hom}^2(z),
	\label{Delz}
\end{equation}
and
\begin{equation}
	\Delta_M^2(M,z) = \frac{\Delta_{\rm vir}(z)}{3}
	\int dC_{\rm vir }(M,z)P(C_{\rm vir}(M,z))\frac{I_2(x_{\rm max})}{[I_1(x_{\rm max})]^2}
	x_{\rm max}^3.
\end{equation}
If one forgets about the formation of DM halos and just assume all DM is distributed homogeneously
in the Universe throughout the whole history, 
the diffuse extragalactic gamma-ray flux would be evaluated by Eq.~(\ref{ext})
with replacing $\Delta^2(z)$ with 1.
Thus $\Delta^2(z)$ represents the enhancement of the gamma-ray flux due to the
structure formation of DM.

In Eq.~(\ref{Delz}), the integration with $M$ has a lower cut ($M_{\rm cut}$),
below which DM cannot form collapsed objects~\cite{Loeb:2005pm}.
The matter density contrast does not grow below the DM free streaming scale.
%Another cut-off scale is provided by the kinetic decoupling of DM.
More important cutoff is provided by dissipation in imperfect fluid at kinetic decoupling of DM.
If DM establishes thermal equilibrium in the early Universe,
it eventually decouples from thermal bath.
In general, the epoch of the end of kinetic equilibrium differs from that of thermal equilibrium.
For example, typical kinetic decoupling temperature $(T_{\rm kd})$ for SUSY WIMPs ranges
$T_{\rm kd} \sim 10~{\rm MeV}$-$1~{\rm GeV}$~\cite{Profumo:2006bv,Bringmann:2006mu}.
Then $M_{\rm cut}$ is estimated as
\begin{equation}
	M_{\rm cut} \sim 10^{-4}M_\odot \left(\frac{\Omega_m h^2}{0.11}\right)
	\left ( \frac{10~{\rm MeV}}{T_{\rm kd}} \right )^3.   \label{Mcut}
\end{equation}
Thus we assume that fraction of DM which is not contained in the halo with mass larger than 
$M_{\rm cut}$ is distributed smoothly.
The fraction of homogeneous component ($f_{\rm hom}$) is given by
\begin{equation}
	f_{\rm hom}(z) =1- \frac{1}{\bar \rho_m} 
	\int _{M_{\rm cut}} dM M \frac{dn(z)}{dM},
\end{equation}
Above the redshift $\sim 50$, the homogeneous component becomes a dominant contribution
to the diffuse extragalactic gamma-rays, as we will see.

\begin{figure}[ht]
 \begin{center}
 \includegraphics[width=0.5\linewidth]{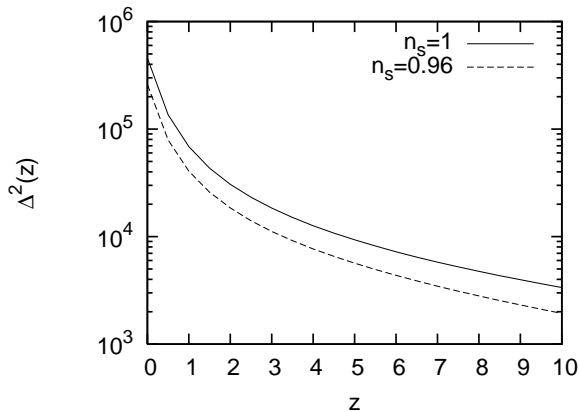} 
 \caption{ Dependence of $\Delta^2(z)$ on the scalar spectral index $n_s$
 	for NFW profile.
 	The upper line corresponds to the scale-ivariant case ($n_s=1$) and 
	the lower line corresponds to the WMAP5 best fit value ($n_s=0.96$), 
	with other cosmological parameters fixed to best-fit values. }
  \label{fig:Delz_ns}
 \end{center}
\end{figure}

Fig.~\ref{fig:Delz_ns} shows a spectral index ($n_s$) dependence on the enhancement factor.
Since the small scale matter power spectrum is affected by scalar spectral index,
the enhancement factor also slightly affected.
In this figure other cosmological parameters are set to be best fit values by WMAP5.
Thus it should be noted that cosmological gamma-ray flux from DM annihilation
depends on cosmological parameters up to numerical factor.

\begin{figure}[ht]
 \begin{center}
 \includegraphics[width=0.5\linewidth]{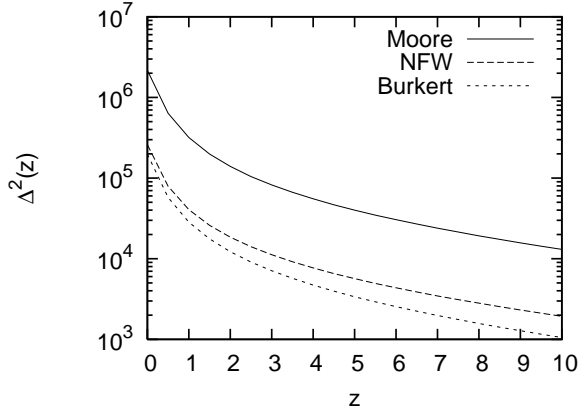} 
 \caption{ Halo density profile dependence of $\Delta^2(z)$.
  The solid, dashed dotted lines correspond to Moore, NFW, Burkert profiles, respectively. }
  \label{fig:Delz_prof}
 \end{center}
\end{figure}

Even larger uncertainty on the gamma-ray flux arises 
due to the lack of the understanding of DM halo profiles.
Fig.~\ref{fig:Delz_prof} shows the dependence of the enhancement factor on DM halo profile.
It is seen that an order-of-magnitude uncertainty exists,
depending on what kind of density profile is taken.
As is expected, steeper profiles predict larger gamma-ray flux.

Although the prediction suffers from these uncertainties, 
it is important to keep in mind that the enhancement factor, $\Delta^2(z)$,
can reach to $\mathcal O ({10^5}$-$10^6$) near $z=0$.
This makes it possible for the DM annihilation-originated gamma-rays
to have significant contribution to diffuse exragalactic gamma-rays.

\begin{figure}[ht]
 \begin{center}
 \includegraphics[width=0.45\linewidth]{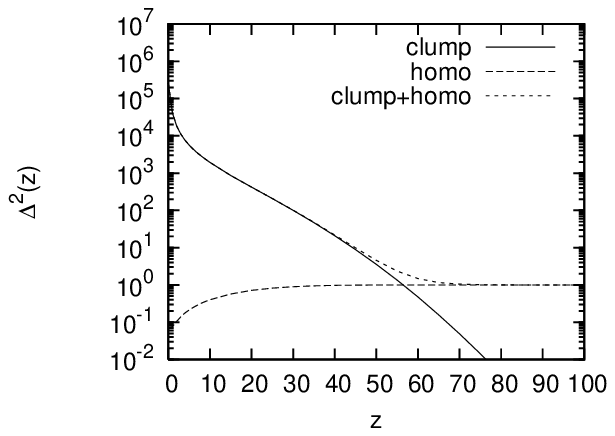} 
 \includegraphics[width=0.45\linewidth]{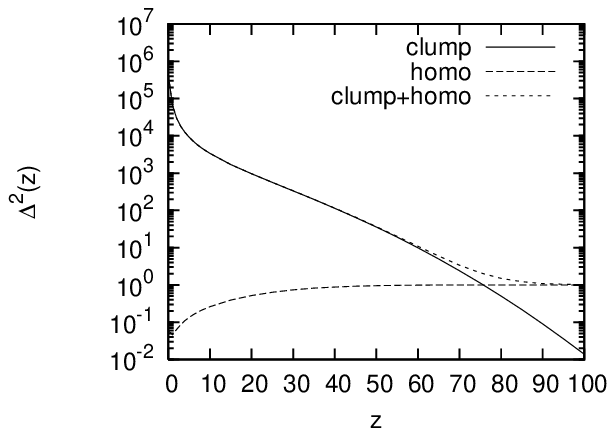} 
 \caption{$\Delta^2(z)$ for $T_{\rm kd}=10$~MeV (top) and 1~GeV (bottom), for NFW profile.}
  \label{fig:Delz_hom}
 \end{center}
\end{figure}

On the other hand, at high redshift the typical mass of DM halos becomes smaller and smaller,
which is even below the smallest possible halo mass (\ref{Mcut}).
This occurs at $z\sim 30$-60 depending on the kinetic decoupling temperature $T_{\rm kd}$.
Before this epoch we can effectively regard DM in the Universe as homogeneously distributed 
without forming clustering objects.
In that regime, DM annihilation rate increases with redshift simply because the physical number density
of DM increases proportional to $(1+z)^3$.
Fig.~\ref{fig:Delz_hom} shows it clearly.
The top (bottom) panel corresponds to $T_{\rm kd}=10$~MeV (1~GeV).
Here we have taken NFW profile.
It is seen that above the redshift 40-60 depending on $T_{\rm kd}$,
the homogeneous component ($f_{\rm hom}$ in Eq.~(\ref{Delz})) becomes more important.

%%%%%%%%%%%%%%%%%%FIGURE%%%%%%%%%%%%%%%%%%%
%%%%%%%%%%%%%%%%%%%%%%%%%%%%%%%%%%%%%%%%%

%%%%%%%%%%%%%%%%%%FIGURE%%%%%%%%%%%%%%%%%%%
%%%%%%%%%%%%%%%%%%%%%%%%%%%%%%%%%%%%%%%%%

%%%%%%%%%%%%%%%%%%FIGURE%%%%%%%%%%%%%%%%%%%
%%%%%%%%%%%%%%%%%%%%%%%%%%%%%%%%%%%%%%%%%

\begin{figure}[t]
 \begin{center}
 \includegraphics[width=0.5\linewidth]{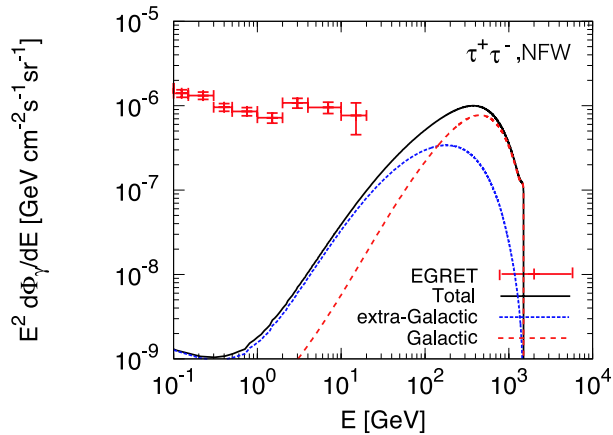} 
 \caption{ Galactic  and extragalactic contributions to gamma-ray flux
  for the case of DM annihilating into $\tau^+\tau^-$ for the model (b) in Table.~\ref{table:model}.
  NFW profile is chosen.}
  \label{fig:gal-ext}
 \end{center}
\end{figure}

Now let us compare the DM-originated gamma-ray flux with the observation by EGRET satellite
\cite{Sreekumar:1997un,Strong:2004ry}.
In order to do so, 
we add the Galactic contribution (\ref{fluxGCpsi}) averaged over the region above the latitude
$|b| >10^\circ$ to the diffuse extragalactic flux given by Eq.~(\ref{ext}) as
\begin{equation}
	\frac{d\Phi_\gamma}{dE}
	= \left[\frac{d\Phi_\gamma}{dE}\right]_{\rm ext}
	+\left \langle \left[\frac{d\Phi_\gamma}{dE}\right]_{\rm Gal} \right \rangle_{|b|>10^\circ}.
\end{equation}
Fig.~\ref{fig:gal-ext} compares the gamma-ray flux from extragalctic contribution
and Galactic component, for the case of DM annihilating into $\tau^+\tau^-$ 
for the model (b) in Table.~\ref{table:model}.
Here NFW halo profile is chosen.
It is seen that both contributions are equally important.
For the case of Moore profile, extragalctic contribution is enhanced by about one order of magnitude,
but the Galactic component almost remains unchanged.

%%%%%%%%%%%%%%%%%%FIGURE%%%%%%%%%%%%%%%%%%%
%%%%%%%%%%%%%%%%%%%%%%%%%%%%%%%%%%%%%%%%%

\begin{figure}[t]
 \begin{center}
 \includegraphics[width=0.45\linewidth]{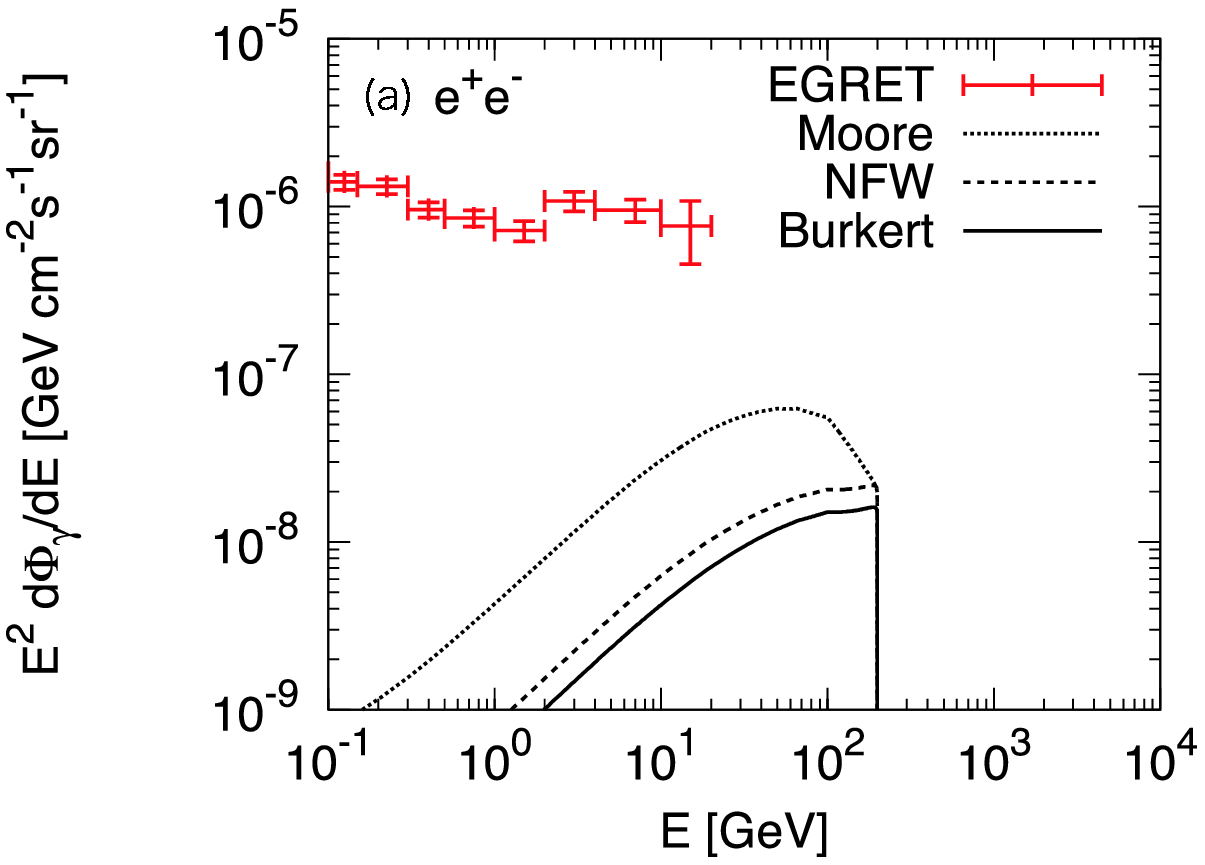} 
 \includegraphics[width=0.45\linewidth]{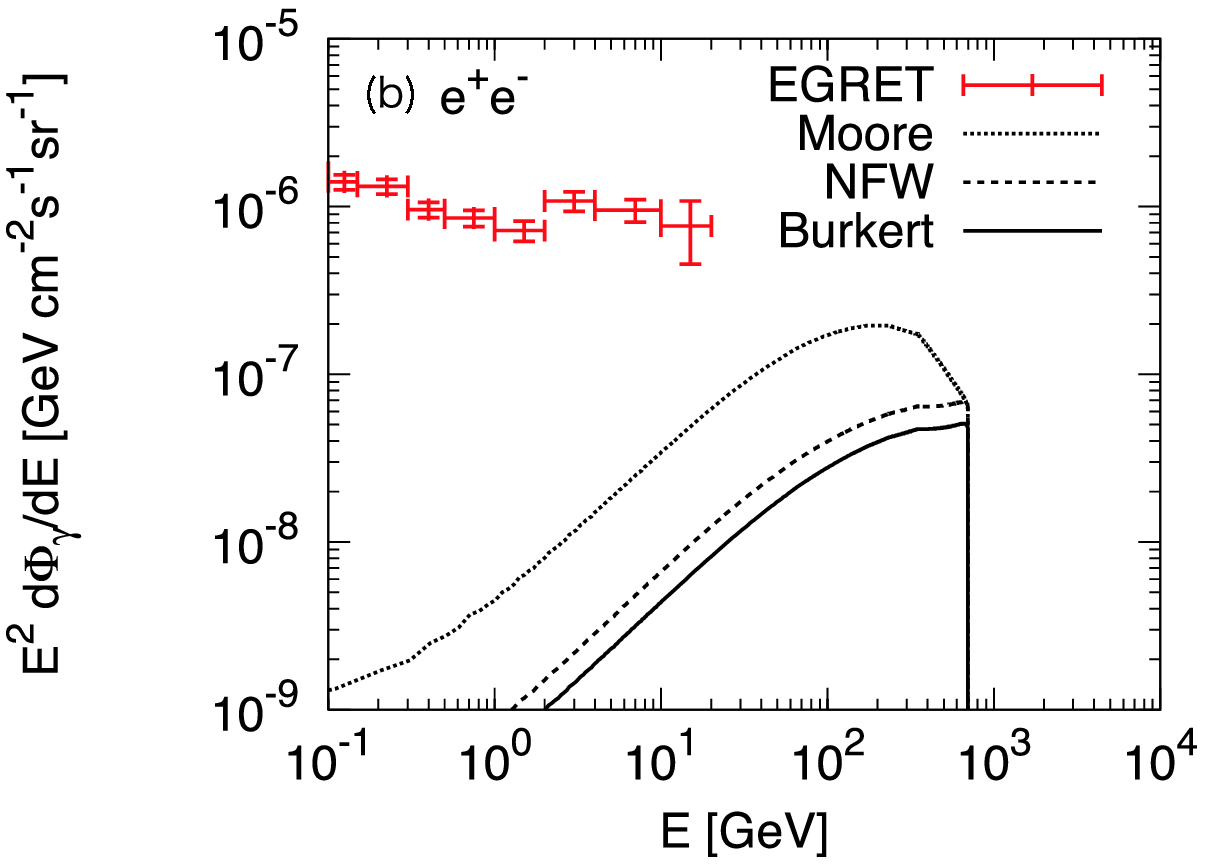} 
 \caption{Diffuse gamma-ray flux from DM annihilation averaged over $|b| >10^\circ$
  and EGRET results for the case of annihilation into $e^+e^-$ for the model (a) (left) and (b) (right) 
  in Table.~\ref{table:model}.}
  \label{fig:ext-e}
 \end{center}
\end{figure}

\begin{figure}[t]
 \begin{center}
 \includegraphics[width=0.45\linewidth]{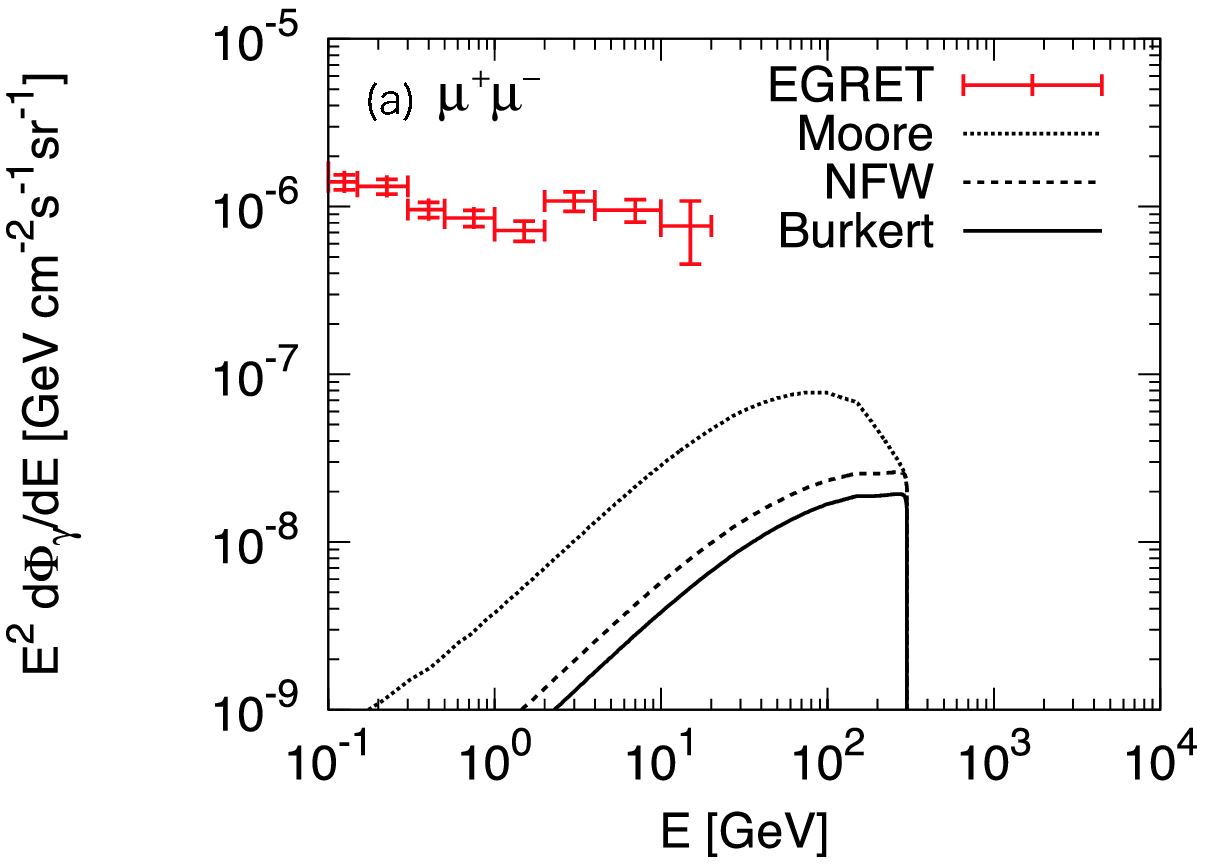} 
 \includegraphics[width=0.45\linewidth]{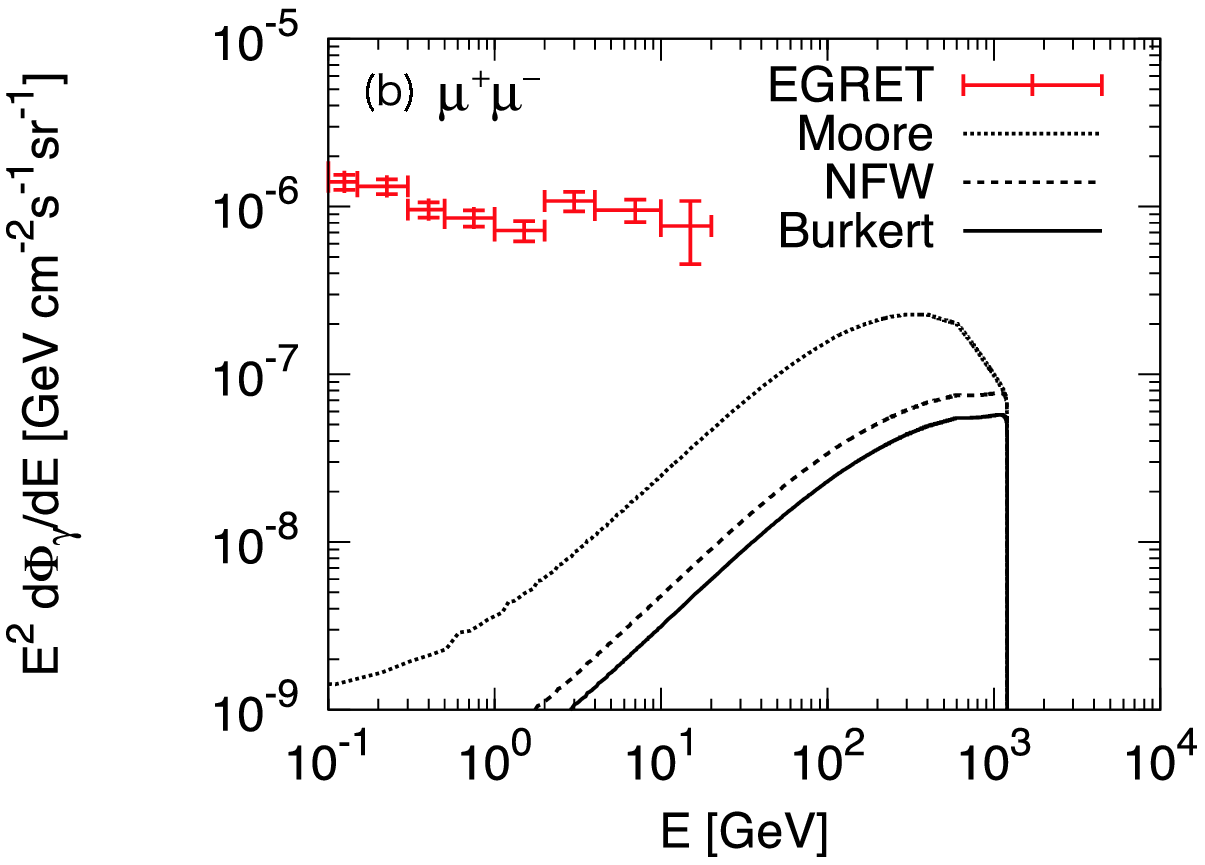} 
 \caption{Same as Fig.~\ref{fig:ext-e}, but for DM annihilating into $\mu^+\mu^-$.}
  \label{fig:ext-mu}
 \end{center}
\end{figure}

\begin{figure}[t]
 \begin{center}
 \includegraphics[width=0.45\linewidth]{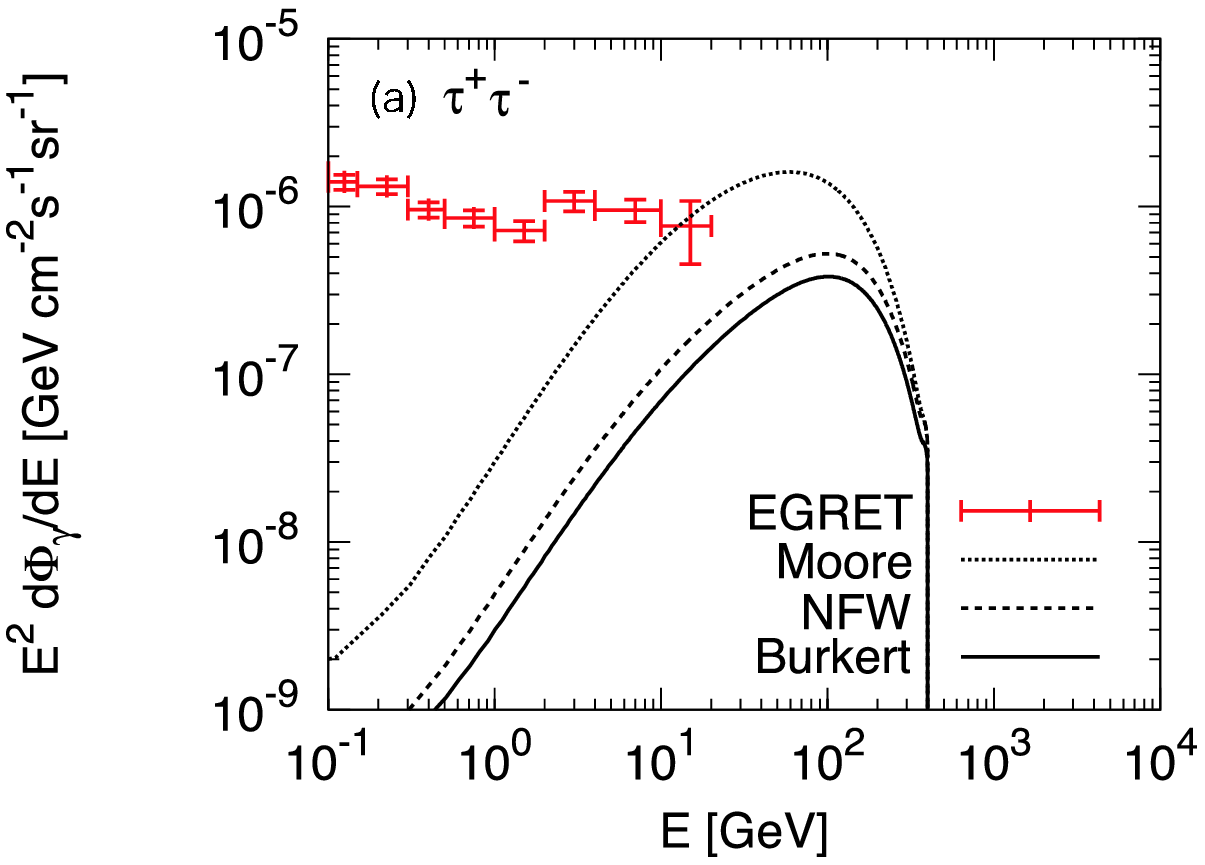} 
 \includegraphics[width=0.45\linewidth]{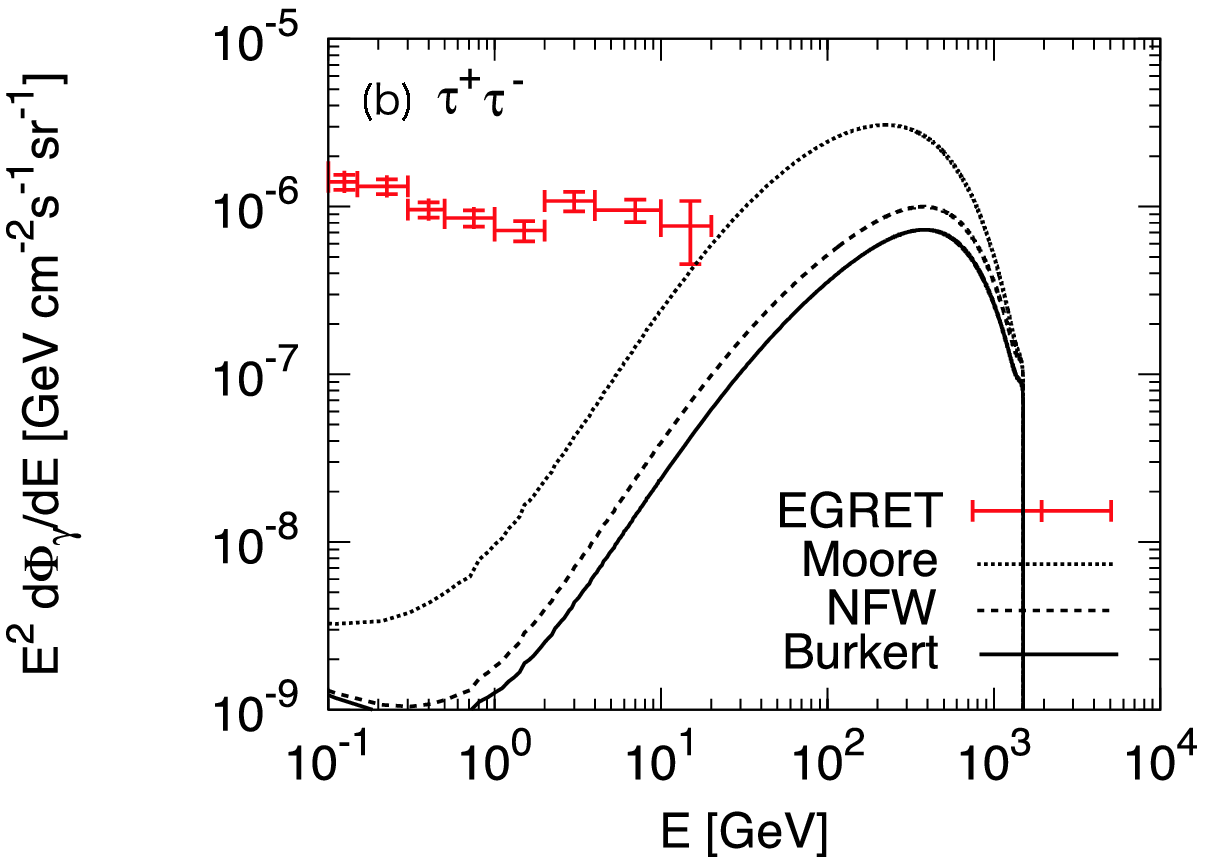} 
 \caption{Same as Fig.~\ref{fig:ext-e}, but for DM annihilating into $\tau^+\tau^-$.}
  \label{fig:ext-tau}
 \end{center}
\end{figure}

\begin{figure}[t]
 \begin{center}
 \includegraphics[width=0.45\linewidth]{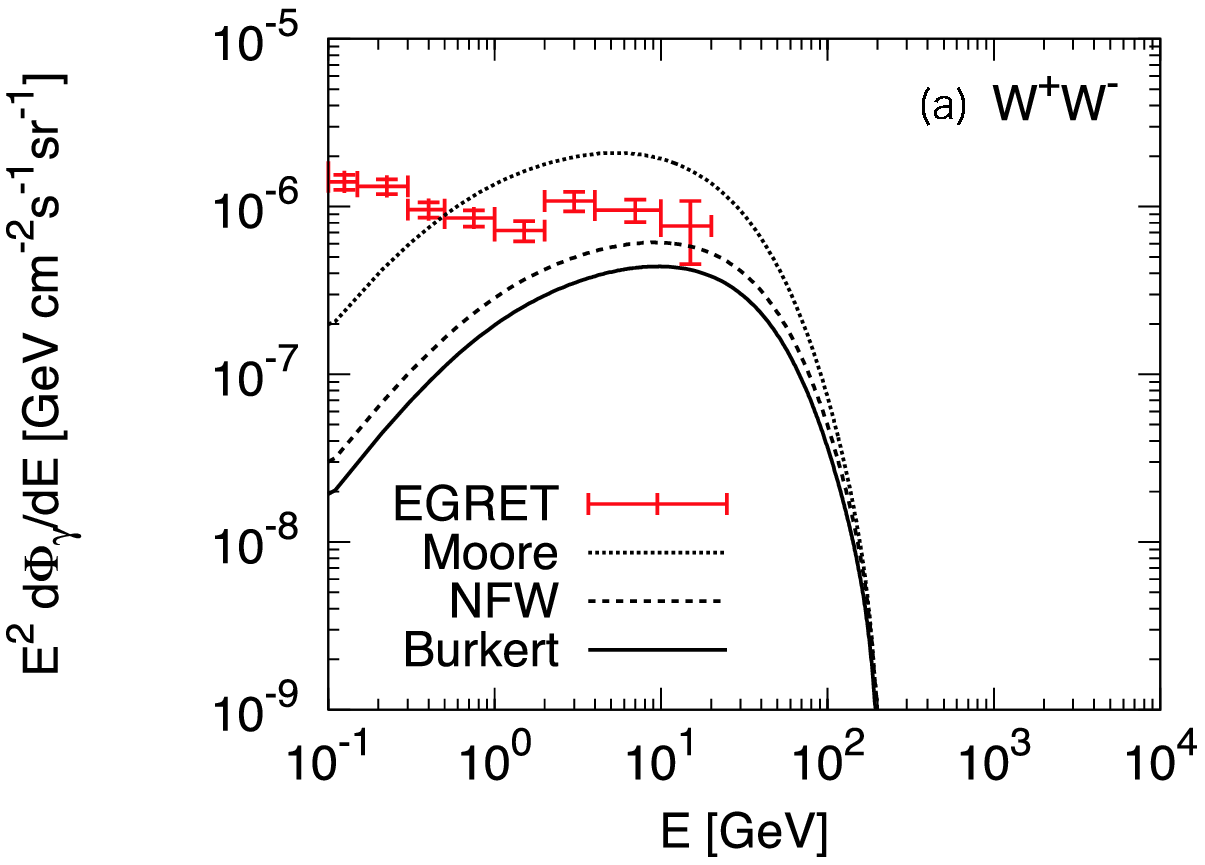} 
 \includegraphics[width=0.45\linewidth]{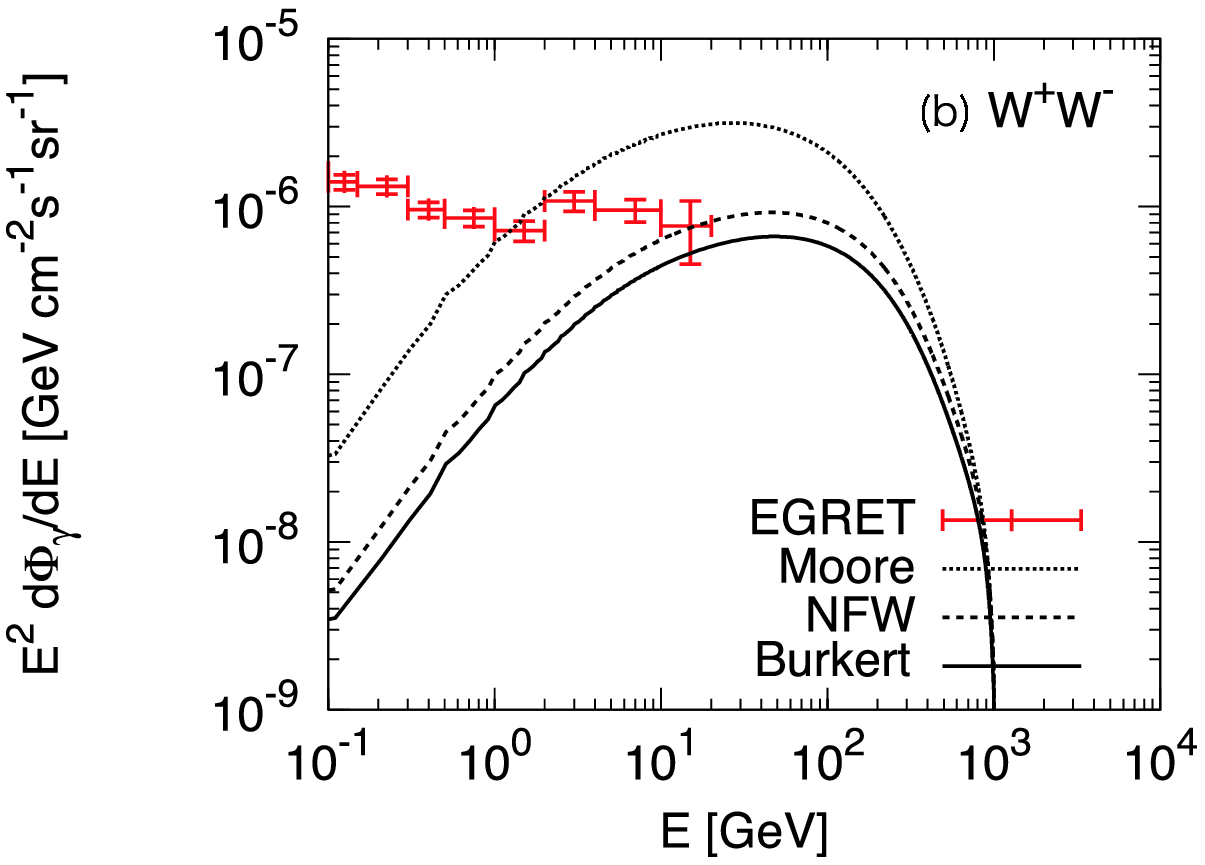} 
 \caption{Same as Fig.~\ref{fig:ext-e}, but for DM annihilating into $W^+W^-$.}
  \label{fig:ext-W}
 \end{center}
\end{figure}

Results are summarized in Figs.~\ref{fig:ext-e}-\ref{fig:ext-W}.
For the case of annihilating into $e^+e^-$ and $\mu^+\mu^-$,
the current data does not conflict with theoretical prediction.
For the case of $\tau^+\tau^-$ and $W^+W^-$, however, 
the prediction may exceed the observed gamma-ray flux for Moore profile,
and NFW or Burkert profile is marginally allowed.
Notice that the diffuse extragalactic gamma-ray flux is not so changed
between the Burkert profile and NFW profile, 
as opposed to the case of gamma-ray from GC (see Fig.~\ref{fig:GC}).
Therefore, for such kind of cored density profile, extragalactic gamma-rays
give more stringent constraint on DM annihilation models than that from GC.
Also note that even if universal DM halos take the form of cuspy structure
such as NFW or Moore profiles, our Galaxy does not necessarily coincide with it.
If our Galaxy is special in the sense that it has moderate DM halo profile like Burkert
or cored isothermal profile, the extragalactic gamma-ray component may be more important
to find or constrain signals of DM annihilation.

%%%%%%%%%%%%%%%%%%FIGURE%%%%%%%%%%%%%%%%%%%
%%%%%%%%%%%%%%%%%%%%%%%%%%%%%%%%%%%%%%%%%

%%%%%%%%%%%%%%%%%%FIGURE%%%%%%%%%%%%%%%%%%%
%%%%%%%%%%%%%%%%%%%%%%%%%%%%%%%%%%%%%%%%%

%%%%%%%%%%%%%%%%%%FIGURE%%%%%%%%%%%%%%%%%%%
%%%%%%%%%%%%%%%%%%%%%%%%%%%%%%%%%%%%%%%%%

%%%%%%%%%%%%%%%%%%FIGURE%%%%%%%%%%%%%%%%%%%
%%%%%%%%%%%%%%%%%%%%%%%%%%%%%%%%%%%%%%%%%

\begin{figure}[t]
 \begin{center}
 \includegraphics[width=0.5\linewidth]{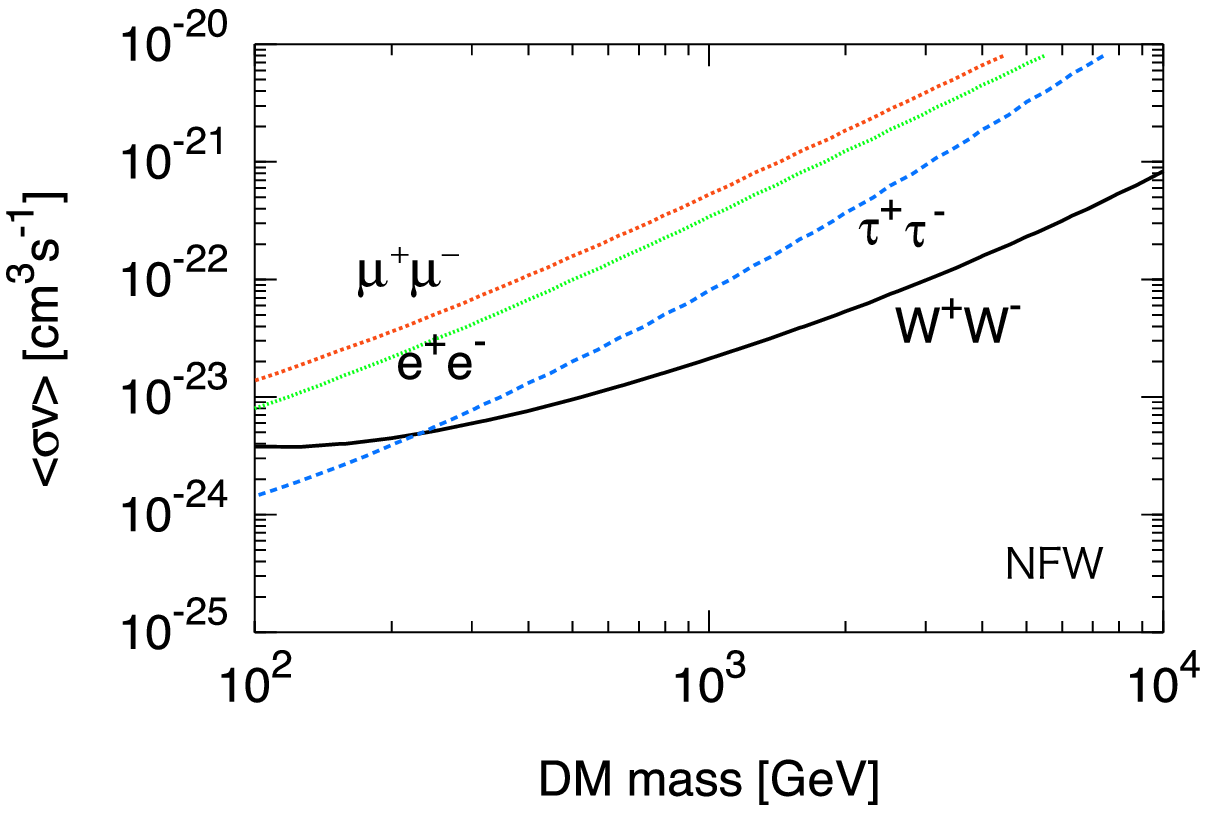} 
 \caption{Constraint on the annihilation cross section from EGRET observations of diffuse extragalactic 
  gamma-ray background for the case of DM annihilating into $e^+e^-,\mu^+\mu^-,\tau^+\tau^-,
  W^+W^-$. NFW halo profile is assumed.}
  \label{fig:const}
 \end{center}
\end{figure}

In Fig.~\ref{fig:const}, constraints on the DM annihilation cross section from the
EGRET observation of diffuse extragalactic gamma-rays are depicted,
for the case of DM annihilating into $e^+e^-,\mu^+\mu^-,\tau^+\tau^-$ and $W^+W^-$,
respectively.
In this figure NFW halo profile is assumed.
The derived constraint is stronger than the constraint from BBN~\cite{Hisano:2008ti}
for the case of leptonic annihilation.
Moreover, as is obvious from previous figures, about one-order of magnitude severer constraint is obtained if Moore profile is adopted and
in this case the constraint becomes more stringent.

It is expected that the Fermi satellite will soon release new data for the diffuse 
gamma-rays, and it will significantly improve the present constraint.

%%%%%%%%%%%%%%%%%%FIGURE%%%%%%%%%%%%%%%%%%%
%%%%%%%%%%%%%%%%%%%%%%%%%%%%%%%%%%%%%%%%%

%%%%%%%%%%%%%%%%%%%%%%%%%%%%%%%%%%%%%%%%%
\section{Conclusions and discussion} \label{sec:conc}
%%%%%%%%%%%%%%%%%%%%%%%%%%%%%%%%%%%%%%%%%

In this paper we have investigated constraints on DM annihilation models
focusing on diffuse extragalactic gamma-ray background,
motivated by DM annihilation scenarios on recent anomalous cosmic positron/electron fluxes.
It is found that it provides comparable or stronger bounds than the gamma-ray observations at GC.
In particular, if the DM halo profile of our Galaxy is moderate, 
the extragalactic gamma-rays are more important than that from GC, for 
constraining or finding the signals of DM annihilation.\footnote{
	Another way to avoid too large gamma-ray flux from GC is to assume that
	DM annihilates into long-lived intermediate particles~\cite{Rothstein:2009pm}.
	Even in such a case, diffuse gamma-rays may provide a useful constraint.
}
If DM contributions to the diffuse extragalactic gamma-ray background is significant,
there may be a chance to extract DM signals from astrophysical contributions
by observing anisotropy~\cite{Ando:2005xg,Cuoco:2007sh} or spectral shape~\cite{Dodelson:2009ih}.

Some comments are in order.
Leptonically annihilating DM would also naturally produce high-energy neutrinos
with similar rate to the charged leptons, if the final states are left-handed.
In this case, the result of Super-Kamiokande~\cite{Desai:2004pq} gives severe constraint on 
DM annihilation models~\cite{Hisano:2008ah},
which is comparable to the gamma-ray constraint (see also Ref.~\cite{Beacom:2006tt}).
Another interesting constraint comes from big-bang nucleosynthesis (BBN).
For DM models with large annihilation cross section,
the annihilation products during the BBN epoch may upset the success of standard BBN prediction,
and it also gives comparable constraints on the annihilation cross section 
for the case of hadronic annihilation modes~\cite{Hisano:2008ti} 
(see also Refs.~\cite{Reno:1987qw,Jedamzik:2004ip} for early attempts).

It is often discussed that non-observations of the excess of cosmic-ray anti-protons
severely constrain DM annihilation models~\cite{Donato:2008jk},
although DM-orinated anti-proton flux suffers from 
large uncertainties regarding the choice of the diffusion zone,
which is largely unknown~\cite{Donato:2003xg}.
If the final state particles of DM annihilation have electric charge, 
they necessarily emit synchrotron radiation due to the Galactic magnetic field.
It was discussed that DM annihilation cross section with standard value
($\langle \sigma v \rangle =3\times 10^{-26}~{\rm cm^{-3}s^{-1}}$)
reproduces the so-called WMAP haze, which is unknown component of the radio emission
around the GC~\cite{Hooper:2007kb}.
According to recent studies~\cite{Borriello:2008gy,Cumberbatch:2009ji},
the constraint is relaxed by two or three orders of magnitude.
Thus it may also give comparable constraints with gamma-rays.
Extragalactic radio emission due to the cosmological DM annihilation
taking into account the effects of DM clustering, similar to the context given in this paper,
was studied in Ref.~\cite{Zhang:2008rs}.

In this study we have assumed that the DM annihilation proceeds via $s$-wave process
and it is time-independent, i.e., $\langle \sigma v \rangle$ is a constant.
In general, however, it could depend on the DM velocity.
Especially, there are some DM models where the annihilation cross section scales as $\propto v^{-1}$
\cite{Hisano:2003ec,ArkaniHamed:2008qn}.
It was discussed that in such case the gamma-ray contribution from the smallest halo
formed in the earliest epoch is most important since the velocity dispersion of DM
in such halos are small and the annihilation rate is significantly enhanced~\cite{Kamionkowski:2008gj}.
BBN may also give stringent constraints since the velocity dispersion of DM in the BBN era is
small.

%%%%%%%%%%%%%%%%%%%%%%%%%%%%%%%%%%%%
\section*{Acknowledgment}
%%%%%%%%%%%%%%%%%%%%%%%%%%%%%%%%%%%%

%%%%%%%%%%%%%%%%%%%%%%%%%%%%%%%%%%%%%%%%%%%%
%\begin{acknowledgements}
%%%%%%%%%%%%%%%%%%%%%%%%%%%%%%%%%%%%%%%%%%%%

K.N. would like to thank the Japan Society for the Promotion of
Science for financial support.  This work is supported by Grant-in-Aid
for Scientific research from the Ministry of Education, Science,
Sports, and Culture (MEXT), Japan, No.\ 14102004 (M.K.), and also by World
Premier International Research Center Initiative, MEXT, Japan.
K.K. is supported in part by STFC grant, PP/D000394/1, EU grant
MRTN-CT-2006-035863, the European Union through the Marie Curie
Research and Training Network ``UniverseNet.''

%%%%%%%%%%%%%%%%%%%%%%%%%%%%%%%%%%%%%%%%%%%%
%\end{acknowledgements}
%%%%%%%%%%%%%%%%%%%%%%%%%%%%%%%%%%%%%%%%%%%%

%%%%%%%%%%%%%%%%%%%%%%%%%%%%%%%%%%%%%%%%%%%%

%%%%%%%%%%%%%%%%%%%%%%%%%%%%%%%%%%%%%%%%%%%%

\end{document}